\documentclass[sigconf,nonacm]{acmart}
\setcopyright{none}
\newcommand{\sysname}{RAGTrace}
\usepackage[capitalise]{cleveref}
\usepackage{xcolor}
\usepackage{booktabs}                  % only used for the table example
\usepackage{amsmath}
\usepackage{amssymb}

\usepackage{enumitem} % 用于\item排版
\newcommand{\REVISE}[1]{#1}
\newcommand{\red}[1]{#1}

\definecolor{retrievalFailureColor}{RGB}{224,145,146}     % #e09192 浅红色
\definecolor{promptVulnerabilityColor}{RGB}{219,192,118}  % #dbc076 暗黄色
\definecolor{generationAnomalyColor}{RGB}{126,176,213}    % #7eb0d5 浅蓝色
\definecolor{standardInconsistencyColor}{RGB}{197,163,208}% #c5a3d0 浅紫色
\definecolor{bleuScoreColor}{rgb}{1.0, 0.0, 0.0} 
\definecolor{embeddingSimilarityColor}{rgb}{0.8, 0.8, 0.20} 
\definecolor{hallucinationIndexColor}{rgb}{0.5, 0.5, 1.0} 
\definecolor{originalConfigColor}{rgb}{0.431, 0.439, 0.475} % rgba(110, 112, 121, 0.7)
\definecolor{ragBeforeColor}{rgb}{0.255, 0.506, 0.851} % rgba(65, 129, 217, 0.7)
\definecolor{ragAfterColor}{rgb}{0.263, 0.686, 0.494} % rgba(67, 175, 126, 0.7)
\definecolor{Blue1}{rgb}{0.494,0.651,0.769}
\definecolor{Gray1}{rgb}{0.827,0.827,0.827}   
\definecolor{DarkGray1}{rgb}{0.588,0.588,0.588}
\definecolor{DarkBlue1}{rgb}{0.247,0.318,0.710}
\definecolor{LightBlue1}{rgb}{0.129,0.588,0.953} 
\definecolor{namedEntityColor}{rgb}{0.129,0.588,0.953}
\definecolor{evidenceSupportedColor}{rgb}{0.298,0.686,0.314}
\definecolor{Orange1}{rgb}{1.0,0.596,0.0}
\definecolor{interestAreaColor}{rgb}{0.7, 0.7, 0.20} 
\definecolor{lessRelevantAreaColor}{rgb}{0.0, 0.0, 1.0}

\begin{document}

\title{\sysname{}: Understanding and Refining Retrieval-Generation Dynamics in Retrieval-Augmented Generation}

\author{Sizhe Cheng}
\orcid{0009-0001-2474-9230}
\authornote{Equal contribution.}
\affiliation{
  \department{Department of Computer Science and Engineering}
  \institution{Southern University of Science and Technology}
  \city{Shenzhen}
  \state{Guangdong}
  \country{China}
}
\email{chengsz2021@mail.sustech.edu.cn}

\author{Jiaping Li}
\orcid{0000-0003-0128-8993}
% \authornote{Equal contribution.}
\authornotemark[1]
\affiliation{
  \department{Department of Computer Science and Engineering}
  \institution{Southern University of Science and Technology}
  \city{Shenzhen}
  \state{Guangdong}
  \country{China}
}
\email{lijp2024@mail.sustech.edu.cn}

\author{Huanchen Wang}
\orcid{0000-0001-9339-1941}
\affiliation{
  \department{Department of Computer Science and Engineering}
  \institution{Southern University of Science and Technology}
  \city{Shenzhen}
  \state{Guangdong}
  \country{China}
}
\affiliation{
  \department{Department of Computer Science}
  \institution{City University of Hong Kong}
  \city{Hong Kong}
  \country{China}
}
\email{wanghc2022@mail.sustech.edu.cn}

\author{Yuxin Ma}
\authornote{Corresponding author.}
\orcid{0000-0003-0484-668X}
\affiliation{
  \department{Department of Computer Science and Engineering}
  \institution{Southern University of Science and Technology}
  \city{Shenzhen}
  \state{Guangdong}
  \country{China}
}
\email{mayx@sustech.edu.cn}

%   ABSTRACT
\begin{abstract}
Retrieval-Augmented Generation (RAG) systems have emerged as a promising solution to enhance large language models (LLMs) by integrating external knowledge retrieval with generative capabilities. While significant advancements have been made in improving retrieval accuracy and response quality, a critical challenge remains that the internal knowledge integration and retrieval-generation interactions in RAG workflows are largely opaque. This paper introduces \sysname{}, an \red{interactive evaluation system} designed to analyze retrieval and generation dynamics in RAG-based workflows. Informed by a comprehensive literature review and expert interviews, the \red{system} supports a multi-level analysis approach, ranging from high-level performance evaluation to fine-grained examination of retrieval relevance, generation fidelity, and cross-component interactions. Unlike conventional evaluation practices that focus on isolated retrieval or generation quality assessments, \sysname{} enables an integrated exploration of retrieval-generation relationships, allowing users to trace knowledge sources and identify potential failure cases. The \red{system's} workflow allows users to build, evaluate, and iterate on retrieval processes tailored to their specific domains of interest. The effectiveness of the \red{system} is demonstrated through case studies and expert evaluations on real-world RAG applications.
\end{abstract}

% arXiv 版本：移除 CCS 分类与描述
% <concept_id>10003120.10003121.10003129</concept_id>
% <concept_desc>Human-centered computing~Interactive systems and tools</concept_desc>
% <concept_id>10010147.10010178.10010179</concept_id>
% <concept_desc>Computing methodologies~Natural language processing</concept_desc>
% <concept_id>10002951.10003317</concept_id>
% <concept_desc>Information systems~Information retrieval</concept_desc>

% \ccsdesc[500]{Human-centered computing~Interactive systems and tools}
% \ccsdesc[500]{Computing methodologies~Natural language processing}
% \ccsdesc[500]{Information systems~Information retrieval}


\begin{teaserfigure}
  \includegraphics[width=\textwidth]{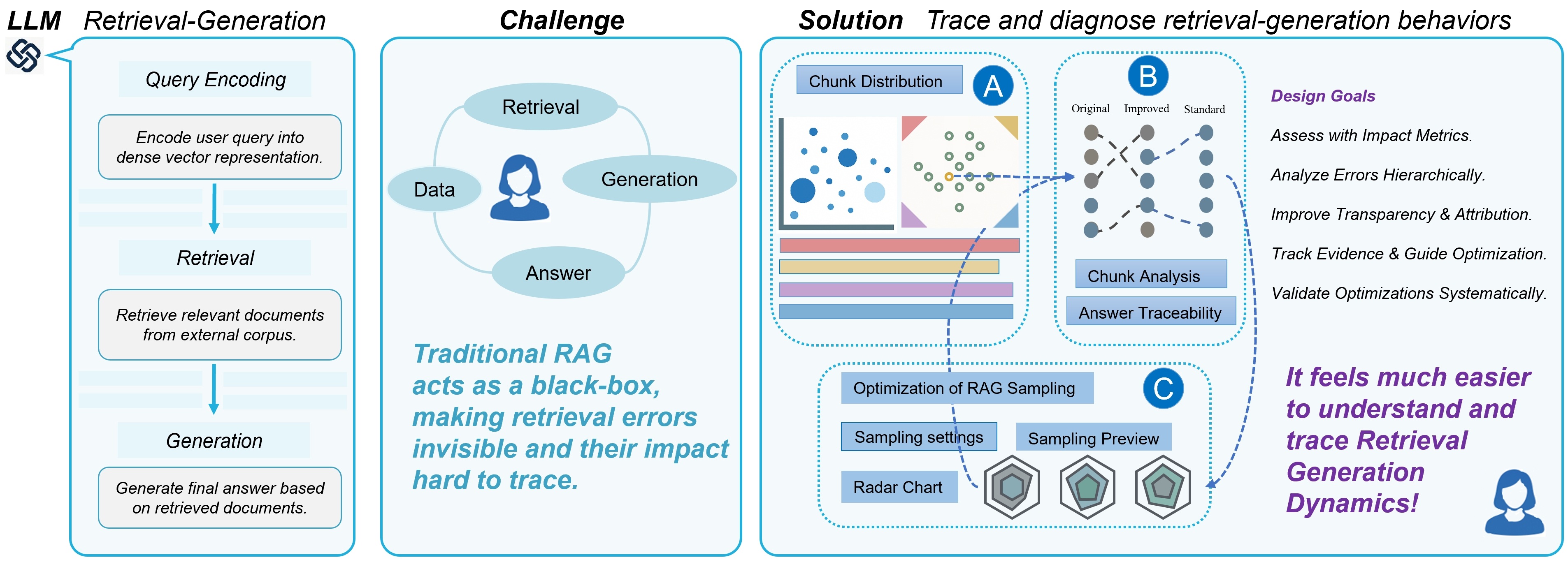}
  \caption{\sysname{} supports interactive refinement of retrieval-augmented generation (RAG) workflows. Users can move beyond passively receiving answers to actively analyzing retrieval-generation dynamics, diagnosing retrieval errors, and steering retrieval strategies for better performance.}
  \label{fig:teaser}
\end{teaserfigure}

\sloppy
\maketitle
\section{Introduction}
As data-driven applications proliferate across AI, scientific computing, and business analytics, the rapid advancement of large language models (LLMs) further amplifies the need for robust evaluation mechanisms to ensure the effectiveness and reliability of visual representations in increasingly complex and AI-driven workflows~\cite{yu2024towards,Kim2024chi}. Despite these impressive achievements, LLMs still face notable challenges in complex, knowledge-intensive tasks that require real-time access to domain-specific information~\cite{fan2024survey,Susnjak2025Automating}. Due to their reliance on static training data, LLMs often struggle with retrieving up-to-date or specialized knowledge~\cite{Raiaan2024Review}, leading to outdated or inaccurate responses. To address this limitation, the Retrieval-Augmented Generation (RAG) paradigm has been introduced, integrating external knowledge sources with generative processes to enable the dynamic retrieval of relevant information before generating model responses~\cite{yu2024evaluation, Soudani2024FineTuning,mansurova2024qa}. Such hybrid approach enhances the factual accuracy, relevance, and overall utility of the generated content, making it particularly valuable for applications in scientific research, business intelligence, and legal documentation~\cite{Sarmah2024HybridRAG}.

Despite the significant advantages of RAG over conventional LLMs, a critical gap remains in the systematic evaluation of RAG-based systems~\cite{wang2023commonsensevis}. Existing research mainly focuses on optimizing retrieval mechanisms and model architectures, yet comprehensive evaluation methodologies remain underdeveloped~\cite{Packowski2025Optimizing, Salemi2024Evaluating}. Specifically, current approaches often assess retrieval precision and generation quality \red{separately}~\cite{Sawarkar2024BlendedRAG}, lacking an integrated system to holistically analyze RAG workflows. \red{Although existing automated evaluation methods primarily emphasize performance metrics and output quality, they often do not explain why RAG behave in certain ways or how to diagnose underlying issues in the retrieval-generation process~\cite{es-etal-2024-ragas, sivasothy2024ragprobeautomatedapproachevaluating}.} \red{Meanwhile, recent advances in the visualization community, aimed at improving model interpretability and shedding light on AI behavior through visual analytics approaches~\cite{wang2023commonsensevis, yan2024knownet, oral2023information}, underscore the effectiveness of visualization-driven techniques for model evaluation and debugging.}
%\red{Recent years have witnessed significant advances in the visualization community's efforts to enhance model interpretability and understand AI behavior through visual analytics~\cite{wang2023commonsensevis, yan2024knownet, oral2023information}, highlighting the effectiveness of visualization-based approaches for model evaluation and debugging.}
Furthermore, the interaction between RAG systems and visualization platforms remains underexplored, despite their crucial role in interpreting AI-generated knowledge and enhancing model transparency \red{for understanding model behavior}~\cite{yan2024knownet}. Visualization is important in interpreting complex data and supporting decision-making~\cite{oral2023information}, yet the absence of a structured evaluation paradigm hinders researchers and practitioners from systematically assessing how well RAG systems retrieve, integrate, and present knowledge. Without a unified evaluation scheme, it remains challenging to refine RAG models for real-world applications and to build user trust through interactive visualization.

%Existing evaluation methods exhibit limitations in scalability and generalizability. Many evaluation studies are domain-specific, limiting their applicability across different tasks and user contexts~\cite{han2024automating}. 
% To address the research gaps mentioned above, we propose a comprehensive interactive evaluation \red{system} for RAG processes by utilizing visualization-based analytical tools. We aim to systematically evaluate both the accuracy of knowledge retrieval and the quality of generated outputs in a cohesive manner, emphasizing their role in AI-enhanced decision-making and user trust. By providing a structured evaluation methodology, we seek to offer a more holistic understanding of RAG's capabilities, supporting its deployment in complex, knowledge-intensive applications that require high levels of interpretability, accuracy, and reliability.

\red{The last decade has seen the community successfully develop evaluation systems to assess AI-driven decision-making systems, ensuring their reliability and interpretability. We believe that the same paradigm is equally relevant to evaluating RAG workflows. Accordingly, this paper introduces an interactive visual evaluation \red{system}, \sysname{}, designed to facilitate the comprehensive evaluation of RAG workflows by analyzing retrieval quality and generative coherence.} \REVISE{Drawing} from an extensive review of relevant literature and collaboration with domain experts in RAG, various analytical tasks have been derived to inform the visualization and interaction design in a hierarchical manner. By employing the ``overview+detail'' scheme, \sysname{} utilizes a multi-level design that exposes the performance of RAG workflows from three perspectives: the analysis of retrieval relevance and redundancy, evaluation of the quality and consistency of generated outputs, and in-depth tracking of error patterns and propagation across retrieval and generation steps. The effectiveness of \sysname{} is demonstrated through case studies on real-world datasets and expert interviews, showcasing its capability to enhance interpretability, diagnose retrieval-generation inconsistencies, and support the optimization of RAG workflows.

In summary, our contributions include:
\begin{enumerate}[leftmargin=*, itemsep=4pt, topsep=0pt]
    \item \textbf{A formative study (N=12)} that captures, through conversational insights, the practices, challenges, and expectations of users in diagnosing, and optimizing RAG outputs.
    \item \textbf{A novel diagnostic methodology} comprising six quantitative and qualitative metrics organized into two categories that systematically evaluate retrieval precision, knowledge transfer efficiency, and generation quality in RAG workflows.
    \item \textbf{\sysname{}}, a comprehensive interactive evaluation \red{system} integrating the metrics to analyze, evaluate, and optimize retrieval-generation interactions, enabling iterative refinement of RAG workflows.
    \item \textbf{Empirical validation through user studies and expert interviews}, demonstrating how \sysname{} helps diagnose retrieval inefficiencies, debug generation errors, and refine retrieval strategies for improved performance.
\end{enumerate}

\section{Background and Related Work}
% Our research addresses the challenges associated with evaluating and interpreting Retrieval-Augmented Generation, where the retrieval and generation components interact in complex ways. Effective \red{interaction} techniques are essential for understanding retrieval quality, document relevance, and response synthesis dynamics. TODO: CHECK HERE
This section reviews existing literature on evaluation methods for LLM and RAG, highlighting current advancements in \red{evaluation} analytics for retrieval processes, generation explainability, and model debugging.

\subsection{Core Process of RAG}
The RAG architecture builds upon a structured pipeline of indexing, retrieval, and generation, which together facilitate the efficient transformation of raw data into contextually grounded outputs~\cite{yu2024evaluation,ragsurvey,barnett2024seven}.
\begin{itemize}[leftmargin=*]
    \item \textbf{Indexing.} Obtain data from dataset and create an index. Specifically, the construction of the data index includes the following steps: 1) Data Preprocessing: Converting raw datasets into a structured text format suitable for retrieval; 2) Chunking: Dividing text into smaller segments~\cite{Coscia2024KnowledgeVIS} to accommodate the context length limitations of LLMs; 3) Embedding and Indexing: Encoding text chunks into vector representations using an embedding model. The resulting vectors, along with their corresponding text, are stored in an index for fast similarity-based retrieval.
    \item \textbf{Retrieval.} Given a user query, the system encodes it into a vector and computes similarity scores against indexed document embeddings. The top-K most relevant chunks are selected as context for the \REVISE{response} generation phase.
    \item \textbf{Generation.} The retrieved context is concatenated with the query and passed to an LLM. Depending on task requirements, the model may either rely solely on retrieved information or incorporate prior knowledge when formulating its response.
\end{itemize}

While these three stages form the foundation of RAG systems, evaluating their effectiveness presents unique challenges, especially in assessing retrieval quality and the impact of retrieved context on the generated response.

\subsection{Evaluation for \REVISE{Retrieval-Augmented Generation}}
% \subsection{Evaluation for Large Language Model}
% \noindent \vspace{0.12mm}\textbf{Evaluation of RAG.} 
The widespread adoption of LLMs across various fields has made the evaluation and analysis of their performance increasingly crucial~\cite{Chang2024ASurvey,Wu2024Survey,Zhao2024Explainability}. Traditional evaluation methods for LLMs primarily rely on automated metrics, such as BLEU~\cite{Papineni2002Proceedings},  ROUGE~\cite{AbuRasheed2024Knowledge}, and human annotations, including expert subjective ratings of generated text~\cite{van2024Field,Bevilacqua2025Automated}. While these approaches assess output quality, they fail to provide deep insights into the underlying mechanisms and decision-making processes of LLMs~\cite{Yang2024Harnessing}. Moreover, due to the inherent ``black-box'' nature of these models~\cite{hassija2024interpreting}, their decision paths remain opaque, making it challenging to explain their behaviors comprehensively.

These challenges become even more pronounced in RAG systems, which couple retrieval and generation modules and thus introduce new layers of complexity to evaluation. While visualization techniques such as causal analysis~\cite{Samarajeewa2024Causal} and uncertainty heatmaps~\cite{Brasoveanu2024Visualizing} can trace error propagation, most tools focus on specific errors rather than providing a comprehensive evaluation environment. \red{Recent automated evaluation frameworks, including RAGAS~\cite{es-etal-2024-ragas}, ARES~\cite{saad-falcon-etal-2024-ares}, and RAGProbe~\cite{sivasothy2024ragprobeautomatedapproachevaluating}, have streamlined RAG workflows assessment through multi-metric integration.} \red{However, these frameworks primarily emphasize metric aggregation, lacking actionable insights for system improvement and interactive visualization for human-in-the-loop RAG optimization.} Additionally, most current approaches evaluate retrieval and generation separately~\cite{Lukas2025Knowledge, Li2025Matching}, overlooking their interdependencies. These limitations highlight the need for evaluation tools that not only provide accurate performance metrics but also support transparent, interpretable, and interactive diagnostics across the entire RAG workflow.

\vspace{-5pt}
\subsection{Interactive Visualization for Evaluating Text Retrieval and Generation}
Interactive visualization has become a vital tool in LLM research to enhance model interpretability, error analysis, and fine-tuning~\cite{Gulsum2022survey,La2023State}, which has significantly improved the usability and transparency of LLMs~\cite{Kahng2025Comparator, Seo2024Seo}. Several interactive systems aid in understanding LLM behaviors from different perspectives. RELIC~\cite{cheng2024relic} evaluates LLM response reliability through self-consistency analysis, allowing users to generate multiple responses and visually identify inconsistencies, particularly hallucinations.
%Its interactive interface decomposes responses into atomic statements and tracks their consistency.
The iScore system~\cite{coscia2024iscore} helps interpret automated scoring of summaries in education, offering parallel visualizations of student responses and model-assigned scores.
%This improves transparency by revealing how text features influence scoring decisions. These systems demonstrate how \red{interactive} visualization enhances trust and interpretability in LLM applications.
These interactive methods have made significant contributions to improving LLM interpretability. However, they are typically tailored for specific tasks (e.g., summarization, consistency checking) or particular models (e.g., ChatGPT self-consistency), which limits their applicability across different contexts. More comprehensive tools such as LLM Comparator~\cite{Kahng2025Comparator, Kahng2024chi}, HaLLMark~\cite{Hoque2024HaLLMark}, and ChainForge~\cite{Arawjo2024chi} offer broader evaluation capabilities by comparing outputs from multiple LLMs, while systems like Aletheia~\cite{Fu2024uist} and WaitGPT~\cite{Xie2024uist} focus on explaining the generation process itself. Despite these advancements, most existing interactive methods analyze individual models in isolation and lack the ability to provide systematic interactive visualizations for complex architectures like RAG, which combine both retrieval and generation processes.

\red{Traditional information retrieval research has established foundational methodologies for understanding user needs and system performance. Ellis et al.~\cite{ellis1989behavioural} proposed a behavioral approach to information retrieval system design, emphasizing the importance of understanding user search patterns and information-seeking behaviors. Building on this behavioral foundation, Chaudhuri et al.~\cite{chaudhuri1998autoadmin} introduced hypothetical analysis techniques that enable quantitative impact assessment of system modifications. As interactive requirements grew to better adapt retrieval systems to user needs~\cite{ingwersen1992information}, Belkin et al.~\cite{belkin1993braque} designed experimental interfaces supporting user interaction by integrating cognitive models with retrieval strategies, while Ahn et al.~\cite{ahn2013adaptive} proposed adaptive visualization techniques to enhance exploratory information retrieval.} These foundational approaches remain relevant in contemporary retrieval-augmented systems, where integrating retrieval with generative models introduces challenges that traditional evaluation methods cannot fully address~\cite{Zhao2024Dense}. Issues like query-document mismatches, vocabulary gaps, and semantic ambiguity~\cite{Cuconasu2024Power} often lead to error propagation in RAG systems, necessitating advanced analysis tools.

Interactive visualization systems help tackle key challenges at different stages of the retrieval-generation pipeline, \red{including retrieval inaccuracies and the opacity of black-box models}. For query formulation, LinkQ~\cite{li2024linkq} provides visual tools to refine queries for knowledge graph question answering. In retrieval evaluation, Angler~\cite{Robertson2023chi} uses coordinated visualizations to detect errors and guide iterative improvements. DeepLens~\cite{Song2023chi} identifies out-of-distribution retrievals that may lead to unreliable generation, while BERT-based similarity mappings~\cite{Julien2024Enhancing} compare retrieved texts and generated outputs to highlight semantic misalignments. VEQA~\cite{VEQA2023} visually explains BERT-based retrieval systems, aiding model interpretability.

Overall, \sysname{} bridges this gap by integrating visualizations of retrieval distributions, semantic similarity mappings, and content attribution. Unlike specialized tools, \sysname{} enables cross-stage error detection, improving transparency and robustness in retrieval-augmented systems.

\section{FORMATIVE INTERVIEWS}
\red{We conducted formative interviews with domain experts and LLM researchers, aiming to understand the black-box nature of RAG systems and uncover unmet evaluation needs.}
\subsection{Participants and Procedure}
We recruited 12 experts through forum posts and word-of-mouth recommendations, consisting of researchers focused on LLMs and RAG (n=4) and domain experts from various fields who frequently use large models (n=8). The participant pool included 8 males and 4 females. 7 participants had over 2 years of experience with LLM, while the remaining five had at least 1 year of experience. Each interview lasted between 30 minutes and 1 hour, and participants were compensated with local currency equivalent to \$15 for their time. \red{Detailed demographic information and background of all participants are provided in \cref{appendix:A}.}

Our participants, labeled E1 through E12, represented diverse expertise backgrounds. E1 to E4 were researchers specializing in LLM with deep understanding of RAG workflows and extensive experience in evaluating the performance of both LLMs and RAG applications. E5 through E12 were domain experts from various fields, each with significant experience using LLMs to address specialized knowledge requirements in their respective domains.

The interviews were designed to uncover practical pain points and research gaps. We began by asking participants about their experiences using and evaluating LLM and RAG workflows, including the strategies they employ to assess system outputs. Next, we explored the limitations of existing evaluation methodologies and the specific challenges they face when interpreting or improving RAG-generated responses. Finally, we discussed their expectations for an ideal evaluation \red{system}, particularly in terms of automation, interpretability, and reliability.

Our interviews followed a semi-structured format divided into three main sections. In the first section, we collected demographic information (gender, occupation) and background data on participants' experience with language models, including preferred models and usage patterns. We then explored their specific use cases, asking questions like ``In which scenarios do you primarily use large language models?'' and ``How frequently do you interact with these systems?'' The second section focused on core research questions regarding RAG systems, including participants' current understanding of RAG technology, their evaluation practices, and their most pressing optimization needs, \red{particularly the improvement of overall performance in users' specific domains}. We inquired about their preferences between holistic versus component-based evaluation approaches and their requirements for interactive evaluation tools. In the final section, participants discussed their expectations for error tracing, automatic optimization across different usage scenarios, and long-term improvement strategies for RAG systems.

During the interviews, we employed contextual inquiry techniques by asking participants to demonstrate their typical workflows when using or evaluating RAG systems. These demonstrations provided valuable insights into practical challenges and workarounds that might not emerge through verbal descriptions alone. All sessions were recorded with permission, transcribed, and subsequently analyzed through qualitative coding. We independently coded the transcripts using an open coding approach, then collaboratively developed a codebook through multiple discussion rounds. The resulting themes were organized into the \red{system} presented in our findings section.

\subsection{Findings}

Based on the interviews, we identified several recurring themes that highlight key challenges in RAG evaluation. These findings are categorized into the following dimensions:

\subsubsection{The Critical Need for Domain-Grounded RAG Solutions.}

Domain expertise integration emerged as a critical requirement, with general-purpose RAG systems failing to meet specialized knowledge needs. Notably, after we provided detailed explanations about RAG to our domain experts, 7 out of 8 expressed strong interest in deploying RAG to enhance knowledge reliability in their respective fields.

\begin{itemize}[leftmargin=*]
    \item \textbf{Knowledge Foundations as a Countermeasure to Model Hallucination.} Hallucination emerged as a pervasive concern among experts working with language models, particularly when applying these systems to specialized domains. The integration of reliable knowledge retrieval serves as a critical anchor to constrain model outputs within factual boundaries. As E3 noted, ``the biggest problem I usually encounter with RAG algorithms is hallucination, sometimes with repetitive answers.'' This phenomenon creates particular challenges in professional contexts where accurate information is paramount. Experts consistently emphasized how retrieval mechanisms must not only fetch information but verify its accuracy before incorporation into responses. Interestingly, even with retrieval-augmentation, models sometimes misinterpret or ignore retrieved information, suggesting that effective hallucination mitigation requires both improved retrieval precision and enhanced reasoning about retrieved content. The interviews revealed a nuanced understanding that hallucinations manifest differently across knowledge domains, requiring tailored detection and prevention strategies rather than one-size-fits-all approaches.
    \item \textbf{Specialized Knowledge Ecosystems Demand Bespoke Retrieval \red{Systems}.} Our interviews revealed a fundamental tension between general-purpose RAG systems and the specialized knowledge requirements of domain experts. Complex fields such as healthcare, engineering, and scientific research operate with distinct epistemological \red{systems} that standard retrieval approaches often fail to adequately capture. E5, working in nuclear engineering, articulated how ``different domains have different focuses'' necessitating ``diversified output methods to address different emphases.'' \red{For instance, nuclear engineering demands high precision in technical documents and zero tolerance for factual errors which could lead to safety risks, while healthcare prioritizes robust prompt fragility and evidence-supported generation anomaly detection.} This insight extends beyond simple keyword matching to encompass domain-specific credibility assessment, contextual relevance, and hierarchical knowledge organization. Particularly revealing was E4's observation regarding the importance of ``credibility/importance of search materials,'' highlighting how domain expertise inherently involves judgments about informational authority that must be encoded within RAG \red{systems}. These findings suggest that truly effective domain-specific RAG systems require not merely access to specialized content, but architectures that embody the evaluation standards and knowledge structures unique to each field of expertise.
\end{itemize}

\subsubsection{Transparency and Evaluation Challenges in RAG Systems.}

The ``black box'' nature of current RAG systems creates substantial barriers to effective assessment and improvement.

\begin{itemize}[leftmargin=*]
    \item \textbf{Information Pathway Visibility: The Black Box Problem in RAG Processes.} A fundamental challenge identified across expert interviews concerns the opacity of information flow between retrieval and generation components. Unlike traditional search systems that present 
    results separately from summaries, RAG systems obscure the relationship between retrieved information and generated responses, creating what several experts described as a  ``black box'' problem. E1 highlighted how ``large models often cannot judge the quality of retrieval results,'' suggesting that failures may cascade undetected through the system. 
    This opacity creates significant diagnostic challenges - when faced with inaccurate outputs, experts struggle to determine whether the retrieval component failed to access relevant information, or if the generation component misinterpreted correctly retrieved content. Our interviews 
    revealed that 6 of 7 domain experts lacked a clear understanding of how RAG retrieves knowledge, and all these experts believed that understanding this mechanism would be meaningful for increasing their trust in model outputs. This lack of transparency fundamentally undermines trust and adoption,  especially in high-stakes domains.
    \item \textbf{Granular Assessment Infrastructure: Beyond Binary Evaluation Paradigms.} Our interviews uncovered a critical gap in evaluation methodology - the lack of nuanced, multi-dimensional assessment \red{systems} tailored to the complex nature of RAG systems. Current evaluation approaches tend toward oversimplified metrics that fail to capture the intricate interplay between retrieval accuracy, information relevance, response coherence, and factual correctness. E5's experience of having ``no relatively quantified feedback'' highlights the reliance on rudimentary methods that are ``very inefficient and cannot provide quantitative results.'' This limitation forces experts to develop ad-hoc evaluation scripts, as noted by E1, or resort to crude comparative techniques. All participants reported experiencing hallucinations in LLMs, with these fabrications potentially undermining trust in model outputs. E3 specifically mentioned encountering ``numerical issues, such as when data mentions hazardous area classifications, the model responds with strange numbers.'' Notably, 10 of 12 participants (E1, E4-E12) stated that hallucination problems lack effective workflows for improvement, making evaluation and enhancement challenging. E5 explained, ``I usually just manually read or check, or use different models to repeat the same work to check result similarity. Both methods are very inefficient and cannot provide quantitative results.'' This sentiment was echoed by E11,  highlighting how current systems lack efficient workflows for identifying and addressing output issues. These insights suggest that advancing RAG systems requires not just improved algorithms but fundamentally new evaluation infrastructures that can decompose performance across multiple dimensions while adapting to the specific requirements of diverse application contexts.
\end{itemize}

\subsubsection{Iterative Development and Collaborative Refinement.}

Participants identified unstructured development processes as a key barrier to effective RAG system improvement.

\begin{itemize}[leftmargin=*]
    \item \textbf{Systematic Performance Evolution: Mapping the Trajectory of System Improvements.} The development of effective RAG systems emerges from our interviews as fundamentally iterative, yet current practices lack structured methodologies for tracking performance evolution across system modifications. Experts described ad-hoc improvement processes characterized by reactive problem-solving rather than systematic optimization. E1's admission of not having ``a systematic iteration method'' where they ``discover a problem and solve it'' reflects a broader pattern of unstructured development processes. This approach creates significant inefficiencies, as improvements in one aspect often produce regressions in others without clear visibility into these trade-offs. E7 noted that ``the biggest challenge is likely the inability to pinpoint problems. Different users need models to solve different problems, so it may be very difficult to develop a systematic iteration method.'' E7 further emphasized the need for ``automatic analysis of problem patterns'' and ``specific case analysis,'' explaining that ``this would be more practical than generalized metrics because it could help developers identify systemic issues rather than addressing symptoms individually.'' These findings suggest that advancing RAG systems requires not just better components but more sophisticated development methodologies that can track multidimensional performance metrics across iterations, enabling developers to understand the holistic impact of incremental changes rather than focusing on isolated improvements.
    \item \textbf{Collaborative Visual Analysis: Enabling Multi-stakeholder System Refinement.} Our interviews revealed strong consensus around the need for interactive visualization tools that transform RAG development from an opaque technical process to a collaborative analytical activity involving diverse stakeholders. Experts consistently expressed enthusiasm for interfaces that would allow real-time parameter adjustment and immediate performance feedback, with E5 noting how such tools ``would significantly improve my efficiency'' for ``rapid iteration and optimization.'' Beyond efficiency gains, these visualization needs reflect a deeper requirement for shared understanding between technical developers and domain experts who may lack programming expertise but possess crucial knowledge about information quality and relevance. E2's anticipation that such tools would be ``very meaningful'' highlights the current absence of accessible interfaces for RAG optimization. E11, from a domain expert perspective, specifically requested ``one-click parameter adjustment functionality for specific optimizations (such as balancing false positives and false negatives),'' emphasizing how domain-specific RAG evaluation requires tailored workflow aligned with specialized knowledge requirements. E9 underscored this point by stating that ``interactive models significantly improve efficiency for trial-and-error and iterative adjustments,'' suggesting that even for users with strong technical backgrounds, the ability to rapidly test hypotheses and observe outcomes remains a critical productivity enhancement. The interviews suggest that effective visualization must go beyond simplistic dashboards to reveal the relationship between system parameters, retrieval patterns, and generation outcomes, enabling intuitive exploration of the complex trade-offs inherent in RAG system design while accommodating stakeholders with varying technical backgrounds.
\end{itemize}

\section{Design Goals}

\REVISE{Based on the themes extracted from our formative interviews and literature review, we derived five design goals to support interpretable and iterative evaluation of RAG workflows. 
The five design goals are as follows:}

\begin{itemize}[leftmargin=*] 
\item \textbf{DG1: Leverage Impact Metrics for Initial Assessment.}
The system should adopt high-level impact metrics that quantify the influence of individual information fragments on the generated outputs. These metrics can serve as early indicators of potential biases and guide further in-depth analysis.
\item \textbf{DG2: Integrate Multi-Level Error Analysis.}  
To comprehensively diagnose failures, the \red{system} must combine global, local, detailed, and relational error analysis. Such an integrated approach will better identify key factors affecting the reliability of the outputs and support targeted refinements. 

\item \textbf{DG3: Enhance Transparency and Attribution.}  
In line with the need for traceability, the system should facilitate the inspection of the generation process by clearly attributing output components to specific retrieved documents or reasoning steps. This goal echoes the importance of explaining and justifying evaluations to support user trust and iterative improvements. 

\item \textbf{DG4: Support Evidence Tracking and Optimization.}  
The \red{system} should allow users to track evidence across different contexts, thereby linking core optimization factors with the underlying retrieval and generation processes. By relating these factors, designers can make more precise adjustments and improve overall result quality. 

\item \textbf{DG5: Enable Systematic Validation of Optimizations.}  
It is crucial to not only identify influencing factors but also to validate and quantify the improvements brought by different optimization strategies. A systematic set of performance metrics will ensure that adjustments are both applicable and generalizable.
\end{itemize}

These design goals collectively form the foundation for an evaluation \red{system} that is both scalable and interpretable, addressing the key challenges identified in our formative interviews.
\begin{figure*}[!ht]
    \centering
    \includegraphics[width=1.00\textwidth]{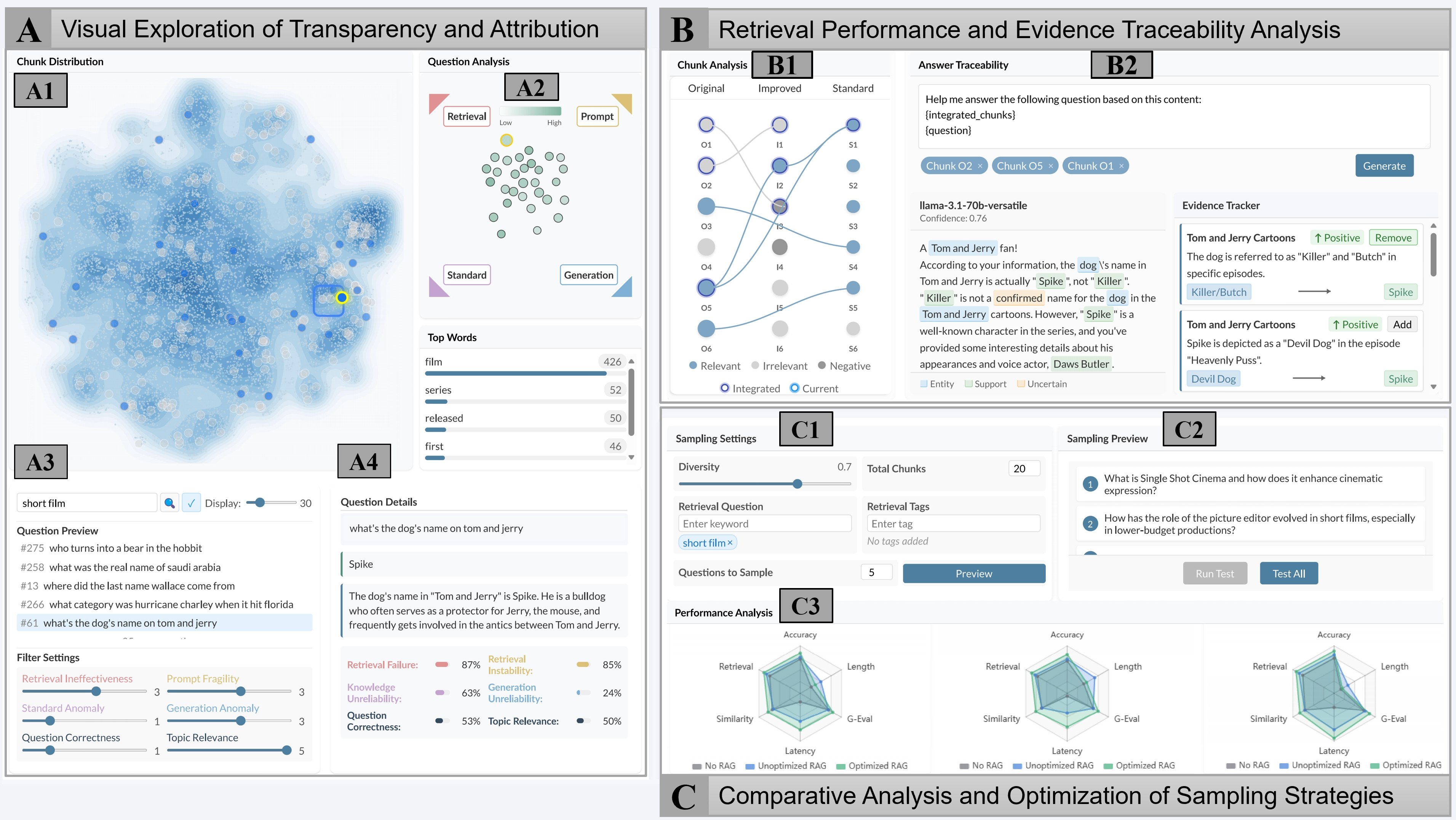}
    \caption{\sysname{} aims to support users refine RAG workflows by interactively analyzing retrieval-generation dynamics, comparing retrieval behaviors, evaluating answer reliability, and optimizing retrieval strategies. In \sysname{}, users examine retrieval behavior through visual analysis and performance metrics (A), evaluate retrieval quality and answer reliability by comparing different retrieval strategies (B), and optimize retrieval configurations by analyzing sampling distributions and performance changes (C).}
    \label{fig:system_main}
\end{figure*}

\section{User Interface}
\red{Based on these design goals, we propose \sysname{}, an interactive evaluation system for tracking and evaluating retrieval-generation dynamics in RAG workflows.}
\red{By following DG2, the entire system employs multi-level error analysis by dividing the interface into three complementary components: (A) visual exploration for global trend analysis and local discrepancy detection, (B) retrieval performance analysis for evidence traceability, and (C) comparative optimization for systematic refinement across different RAG workflow stages.}

\subsection{Visual Exploration of Transparency and Attribution}
In this component shown in~\cref{fig:system_main} (A), a heatmap and Force-directed Graph metrics, along with corresponding visualization modules, are utilized to provide a detailed analysis of chunk distribution and question classification. These modules also offer comprehensive data regarding the questions to support in-depth understanding of answer generation after retrieval, thereby enhancing the transparency and traceability of the RAG workflow. Additionally, manual search capabilities for specific questions are provided to further explore the details associated with each question.

\subsubsection{Heatmap for Chunks Distribution Analysis.} 
Drawing from established visualization techniques for spatial data analysis, the chunk distribution is designed to show to key factors: chunk density and uniformity. \red{In this module, the chunks from the knowledge base are encoded as small white dots positioned based on their semantic embeddings using dimensionality reduction (detailed in Section ~\ref{sec:data_processing}).} Users can perceive the concentration of retrieved chunks through the background intensity in the heatmap; darker backgrounds indicate higher chunk density while uniform background color intensity suggests evenly distributed chunks. White points on the heatmap represent individual chunks, with a global limit of 20,000 chunks to prevent performance issues. When users hover over individual question or chunk, relevant document metadata is immediately displayed.

As illustrated in ~\cref{fig:system_main} (A1), the chunk distribution is displayed in a heatmap. The intensity of the background color denotes chunk density, with darker regions reflecting higher chunk concentrations. For questions, the yellow color highlights mark areas of interest, while blue represents less relevant or unimportant chunks. An interactive zooming feature enables detailed examination of specific chunk regions, with synchronized zooming across all visualizations to facilitate direct comparison. 

The heatmap is divided into a grid of adjustable-sized square cells to facilitate granularity control based on their analytical needs. When a question is selected, the system automatically highlights the corresponding cell and performs a named entity extraction and thematic analysis on all chunks in the cell. This generates a concise set of topic keywords displayed in a dedicated summary view in the lower right corner.

\subsubsection{Force-directed Graph Metrics for Question Analysis.} \red{Aligned with DG1, the Force-directed Graph utilizes composite performance metrics for node positioning and color coding, enabling rapid identification of high-impact failure patterns.} Understanding the failure patterns in RAG is crucial for diagnosing retrieval effectiveness and enhancing generation quality. To achieve this, we utilize a Force-directed Graph visualization, ~\cref{fig:system_main} (A2), that maps questions based on their distance from an ideal retrieval and generation outcome.

The nodes in the Force-directed Graph represent four different question types, with each vertex applying an attractive force to question nodes that exhibit similar characteristics. \red{The question nodes represent test questions filtered and selected through the search interface,~\cref{fig:system_main} (A3)}, displaying only questions of interest to the user. Question nodes simultaneously repel each other to prevent overcrowding and maintain visual clarity. Users can hover over any node to display detailed evaluation information, including specific metrics and contextual details. The identified failure types are categorized as follows:
\begin{itemize}[leftmargin=*]
    \item \textcolor{retrievalFailureColor}{\textbf{Retrieval Failure}}: Highlighted in \textcolor{retrievalFailureColor}{red}, these problems arise when critical chunks are not retrieved, often due to excessive dispersion in retrieval results. High retrieval dispersion is visualized through cosine similarity distributions, where a wider spread indicates a lack of focus.
    \item \textcolor{promptVulnerabilityColor}{\textbf{Prompt Vulnerability}}: Highlighted in \textcolor{promptVulnerabilityColor}{yellow}, this category captures errors stemming from ambiguous prompts that lead to multiple interpretations. The distance between different possible responses serves as a measure of ambiguity, with larger distances indicating greater uncertainty. 
    \item \textcolor{generationAnomalyColor}{\textbf{Generation Anomalies}}: Highlighted in \textcolor{generationAnomalyColor}{blue}, these errors occur during response generation and are characterized by deviations in content attribution. Metrics such as hallucination rates and reference errors (incorrectly citing retrieved chunks) are used to quantify these anomalies.
    \item \textcolor{standardInconsistencyColor}{\textbf{Standard Inconsistencies}}: Highlighted in \textcolor{standardInconsistencyColor}{purple}, these problems emerge from outdated or incorrect information sources, insufficient knowledge base coverage, or errors in the evaluation dataset. Outliers in this category include ambiguous questions or flawed reference answers.
\end{itemize}
The magnitude of each problem directly influences the strength of attraction exerted on corresponding question nodes, where more significant problems create stronger attractive forces and cause question nodes to cluster around the most problematic nodes. The repulsive forces between question nodes promote optimal spacing for interactive exploration, minimize visual clutter, and preserve the visibility of relationships. 

\subsubsection{Interest-driven Question Search.} Providing manual search for related questions is essential for refining query exploration and improving retrieval accuracy. This module,\red{~\cref{fig:system_main}~(A3)} provides an interactive search interface for locating and filtering questions based on specific criteria. The interface includes:
\begin{itemize}[leftmargin=*]
    \item A search bar for direct query input;
    \item A suggested question list displaying example queries;
    \item A result ranking mechanism that sorts retrieved questions by similarity scores and hallucination values.
\end{itemize}

Our system supports four Granular Diagnostic Metrics (\red{\textbf{Retrieval Failure Value}}, \red{\textbf{Prompt Fragility Value}}, \red{\textbf{Standard Hallucination Value}}, \red{\textbf{Generation Anomaly Value}}) and two Composite Performance Metrics (\red{\textbf{Question Correctness}} based on BLEU/ROUGE, and \red{\textbf{Topic Relevance}} based on embeddings). Users can select preset weighting configurations (e.g., prioritizing high hallucination, similarity matching, or improvement potential) and adjust question display proportions via an integrated slider. The question preview pane is bidirectionally bound with nodes in the heatmap and the Force-directed Graph visualization, enabling synchronized highlighting and consistent information display when selecting elements. \red{All questions found in the search interface are encoded as highlighted dots in the heatmap and as question nodes in the Force-directed Graph visualization, ensuring coordinated illustration across all three visualization modules.}

\subsubsection{Evaluation Indicators for Question Details.} To provide an intuitive representation of question evaluation metrics, we employ a progress bar visualization, \red{~\cref{fig:system_main}~(A4)}, to depict \textbf{Granular Diagnostic Metrics} used during the problem localization phase. These indicators are visualized as progress bars, each corresponding to one metric. The length of each bar represents the magnitude of the respective measure, while distinct colors help differentiate between them. 
% This view presents both the model-generated answer and the ground truth answer (i.e., the standard reference from the test dataset) in parallel to enable direct comparison. In addition, the module provides essential metadata, including the Question ID, Question Type, and the number of Related Chunks retrieved during the process, offering a comprehensive overview.
This view displays the model-generated answer and the ground truth answer (i.e., the standard reference from the test dataset) side by side for direct comparison. It also provides key metadata, including Question ID, Question Type, and the number of Related Chunks retrieved, to support a comprehensive overview.

% The entire module is dynamically coordinated with other views. When selecting a different question in either the heatmap or Force-directed Graph visualization, this detailed view automatically updates its content to reflect the evaluation metrics, answers, and metadata of the newly selected question.

The module is dynamically linked with the other views. When a different question is selected in the heatmap or the Force-directed Graph view, the detailed view automatically updates its content to reflect the evaluation metrics, answers, and metadata for that newly selected question.

\subsection{Retrieval Performance and Evidence Traceability Analysis}
A comprehensive evaluation of retrieval quality and evidence traceability is crucial for assessing the reliability of generated responses, going beyond mere parameter metrics and statistical data.
\red{This component fulfills the requirements in DG3 by implementing comprehensive transparency and attribution mechanisms through both retrieval flow visualization and evidence traceability analysis, enabling clear mapping of outputs to their retrieval sources and systematic tracking of information provenance throughout the generation process.} \red{The features are fulfilled by adopting efficient comparative analysis of original, optimized, and reference retrievals.}

\subsubsection{\red{Chunk-Relink Graph} for Retrieval Flow Analysis.}  
A comprehensive examination of retrieved document chunks is essential for understanding retrieval effectiveness and relevance distribution. This module enables comparative analysis across different retrieval strategies, i.e., original, optimized, and reference retrievals, which allows users to assess ranking shifts, content variations, and retrieval consistency within the main workspace,~\cref{fig:system_main} (B1). By visualizing retrieval flow using the \red{Chunk-Relink Graph} diagram, how retrieved chunks evolve across different retrieval configurations can be highlighted.
\begin{itemize}[leftmargin=*]
\item  \textbf{Visual Encoding.} To directly convey retrieval relevance and facilitate comparative analysis, a categorical color scheme is employed to distinguish chunk significance and retrieval status:  
\end{itemize}

\begin{figure}[!b]
	\centering	
	\includegraphics[width=1.0\columnwidth]{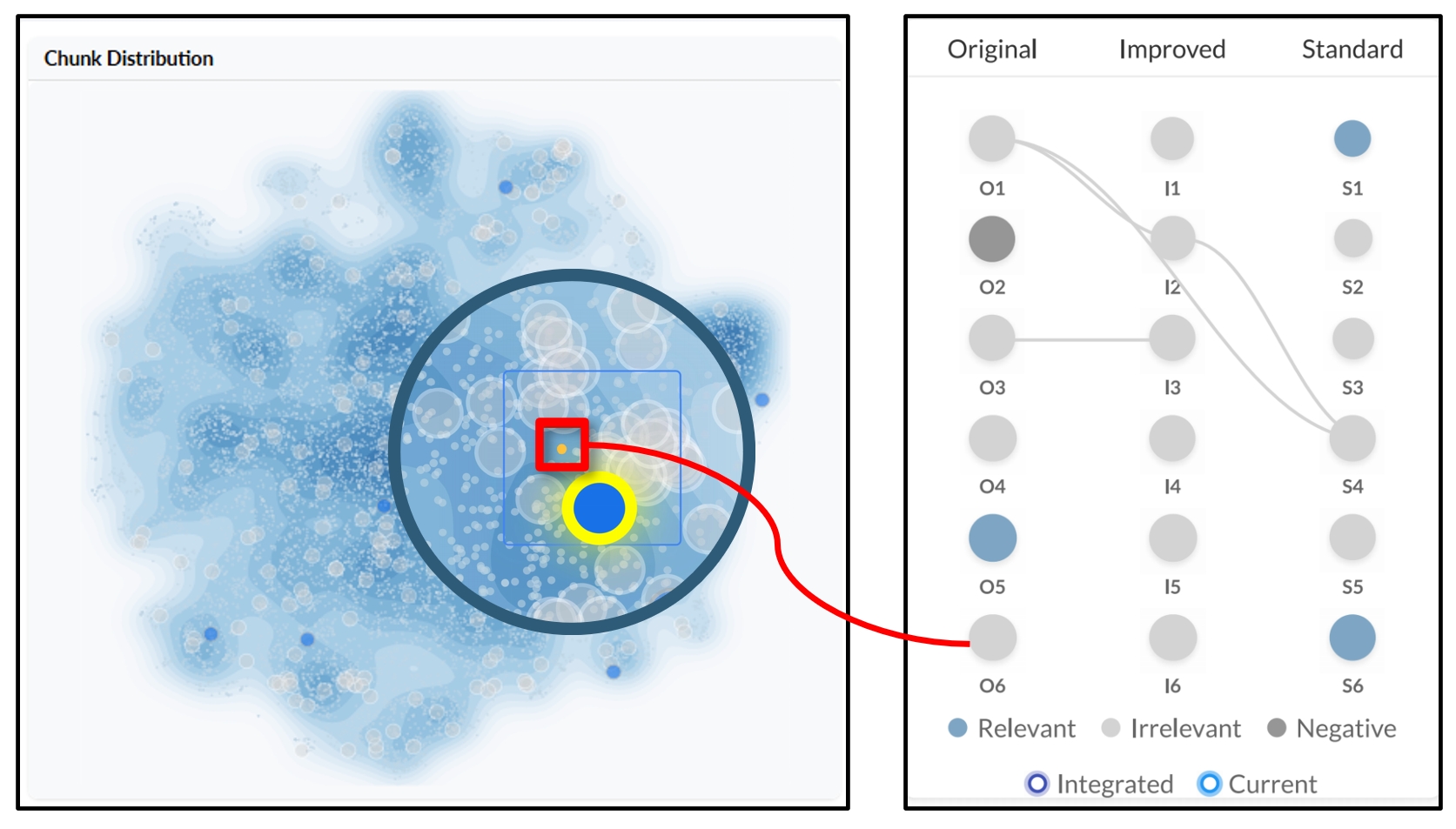}
    \caption{The small dots on the left correspond to the circular nodes in the \red{Chunk-Relink Graph}, both representing a chunk in the RAG knowledge base.}
    \label{fig:chunkAB}
\end{figure}

\begin{itemize}
    \item \textcolor{Blue1}{Blue} – Relevant chunks, strongly contributing to response generation;
    \item \textcolor{Gray1}{Gray} – Irrelevant chunks, providing little contextual information;
    \item \textcolor{DarkGray1}{Dark Gray} – Negative information chunks, which may introduce noise or inconsistencies;
    \item \textcolor{DarkBlue1}{Dark Blue} rings – Integrated chunks that have been merged or aggregated from multiple retrieval sources;
    \item \textcolor{LightBlue1}{Blue} rings – Currently selected chunks under analysis or user interaction.
\end{itemize}
The size of each chunk node encodes its semantic similarity with the query, where larger nodes represent higher cosine similarity in the embedding space between the chunk and the query statement used for retrieval. \red{Guided by DG4, the system implements a synchronized selection mechanism and maintains visual consistency across components. This enables evidence tracking and optimization through coordinated interactions.} This visual cue provides an immediate indication of retrieval relevance independent of the color-based categorization. When identical chunks are retrieved through multiple retrieval methodologies, connecting lines are drawn between these nodes. In addition, selecting or interacting with any chunk automatically highlights all identical chunks across different retrieval strategies.

\subsubsection{Generation Confidence and Evidence Traceability.}  
Ensuring the reliability and traceability of generated answers is crucial for evaluating RAG models, as illustrated in ~\cref{fig:system_main} (B2). We employ a structured visualization approach that highlights confidence levels and evidence sources within responses.
\begin{itemize}[leftmargin=*]
    \item \textbf{Confidence-Based Annotation.} We use a categorical highlight scheme to indicate content reliability: \textcolor{namedEntityColor}{blue} for named entities, \textcolor{evidenceSupportedColor}{green} for well-supported information, and \textcolor{Orange1}{orange} for uncertain or weakly backed content. This visual differentiation helps users quickly assess response robustness.
    \item \textbf{Interactive Evidence Traceability.} Users can modify evidence sources in real time, triggering dynamic updates to highlights and enabling direct comparison of different \red{evidence chunk sets used to generate the target answer}. A directed graph visualization links named entities to supporting evidence chunks, offering an intuitive representation of information provenance.
\end{itemize}

\subsection{Comparative Analysis and Optimization of RAG Sampling Strategies}

\red{Following DG5, we design a comprehensive resampling evaluation component that enables systematic validation through comparative analysis of different RAG configurations.} Building upon the transparency analysis from previous components, such optimization scheme provides structured interfaces for parameter adjustment, real-time preview, and performance tracking across multiple retrieval strategies,~\cref{fig:system_main} (C).

\subsubsection{Sampling Settings.} 
This \red{module} provides a structured interface for configuring key sampling parameters that directly influence retrieval diversity, response consistency, and overall generation quality (C1). The configurable parameters include: \textit{Diversity} (controlling generation randomness to balance creativity and reliability), \textit{Number of Chunks} (determining the context volume with trade-offs between information completeness and redundancy), \textit{Retrieval Keywords and Tags} (refining retrieval focus through term prioritization and metadata filtering), and \textit{Number of Questions to Sample} (defining evaluation scope with considerations for statistical robustness versus computational efficiency). As depicted in ~\cref{fig:system_main} (C1), these parameters are presented within an interactive configuration panel with real-time effects visualization and parameter sensitivity analysis for optimizing RAG strategies.

\subsubsection{Sampling Preview.} 
Building on established methods for evaluating RAG strategies, this \red{module} offers an direct preview of sampling outcomes, ~\cref{fig:system_main} (C2).

\subsubsection{Radar Chart Visualization.} As shown in~\cref{fig:system_main} (C3), this \red{module} visualizes the performance of different configurations using Radar Charts, where each chart represents a single question. The charts are designed to simultaneously display the performance of the original configuration, RAG before optimization, and RAG after optimization.

\begin{itemize}[leftmargin=*]
    \item \textit{Color-coded Stages:} In each Radar Chart, the performance metrics for the original configuration are depicted in \textcolor{originalConfigColor}{gray}, those for RAG before optimization in  \textcolor{ragBeforeColor}{blue}, and those for RAG after optimization in \textcolor{ragAfterColor}{green}, as a superposed manner.
    \item \textit{Interactive Highlighting:} Users can click on any segment of a Radar Chart to view detailed numerical performance values, facilitating an in-depth analysis of specific metrics. 
\end{itemize}

% As shown in ~\cref{fig:system_main} (\textbf{C3}), each Radar Chart offers a comprehensive view of the performance for a particular question across the three configurations.

\section{Implementation}
In this section, we introduce the technical implementation details of \sysname{}, including our evaluation metrics and the data processing pipeline that powers our visualization and analysis capabilities.

\subsection{Evaluation Metrics}
To comprehensively assess RAG systems, we categorize the evaluation metrics into two complementary classes: \textbf{Composite Performance Metrics} for end-to-end performance measurement, and \textbf{Granular Diagnostic Metrics} for modular component analysis. While composite metrics like BLEU, ROUGE and GPTScore provide efficient, standardized evaluation of overall output quality, their reliance on system-level assessment often masks underlying failures in specific components of the RAG workflow~\cite{Sakar2025rag}. 

To investigate the scientific validity and necessity of granular metrics, we conducted an extensive literature review examining failure modes in RAG systems. Recent literature has identified several key failure patterns that require targeted evaluation techniques:

\begin{itemize}[leftmargin=*] 
\item \textbf{Retrieval Failure} occurs when the retrieval module does not capture key evidence or exhibits poor result distribution. Ru et al.~\cite{ru2024ragchecker} introduce fact-level metrics like ``Claim Recall'' to measure factual coverage, while Saad-Falcon et al.~\cite{saad-falcon-etal-2024-ares} propose ARES to assess context relevance. Chen et al.~\cite{chen2024benchmarking} benchmark LLMs in RAG settings, highlighting retriever performance as a critical factor.

To quantify retrieval failures, we define the Retrieval Failure Value (\(\mathcal{R}_{fail}\)), which evaluates both the hit rate on key document chunks and the entropy of the retrieval distribution:
\begin{equation}
\mathcal{R}_{fail} = \alpha \frac{\sum_{c_k \in C_{\text{gold}}} \mathcal{T}_{\theta}(\text{sim}(c_k, C_{\text{ret}}))}{|C_{\text{gold}}|} + \beta \frac{1}{n} \sum_{i=1}^{n} \text{Entropy}\bigl(\text{sim}(c_i, C_{\text{ret}})\bigr)
\end{equation}

where \( \text{sim}(c_i, C_{\text{ret}}) \) is the cosine similarity between chunk \( c_i \) and retrieved chunks, and \(C_{\text{gold}}\) is the set of ground truth chunks. The threshold function \( \mathcal{T}_{\theta}(x) \) indicates retrieval success when similarity exceeds \( \theta \). Entropy is computed as \(\text{Entropy}(X) = -\sum_{x \in X} p(x) \log p(x)\), capturing retrieval concentration.

\item \textbf{Prompt Fragility} occurs when ambiguous prompts lead to inconsistent outputs. Zhang et al.~\cite{zhang-etal-2024-clamber} and Zhuo et al.~\cite{zhuo-etal-2024-prosa} propose benchmarks and metrics to assess model robustness, showing that larger models perform better but remain sensitive to prompt phrasing. To measure fragility, we define the Prompt Fragility Value, which captures retrieval divergence across \( m \) prompts, each retrieving \( n \) chunks:
\begin{equation}
C_{\text{sem}} = \frac{1}{n(n-1)m} \sum_{i=1}^{n} \sum_{j \neq i}^{n} \sum_{k=1}^{m} \max_{l \in [1,m]} \bigl( \text{sim}(d_{i,k}, d_{j,l}) \bigr)
\end{equation}

where \( d_{ik} \) is the \(i\)-th chunk from the \(k\)-th prompt, and \( \text{sim}(d_{ik}, d_{jl}) \) measures semantic similarity between chunks retrieved by different prompt variations.

\item \textbf{Generation Anomaly} occurs when generated answers contain factual errors or inappropriate content. Yue et al.~\cite{yue-etal-2023-automatic} propose methods for evaluating attribution correctness, while Datta et al.~\cite{pmlr-v176-datta22a} introduce the TruLens system to detect hallucinations. Recent approaches focus on automatic fact-checking and citation evaluation.

To measure generation anomalies, we define the Generation Anomaly Value, which combines two components: the model's self-reported confidence and the proportion of erroneous citations:
\begin{equation}
\mathcal{A}_{gen} = \alpha \cdot \left(\text{Mean Confidence}\right) + \beta \cdot \left(\text{Error Chunk Ratio}\right)
\end{equation}

where \(\alpha\) and \(\beta\) are adjustable parameters, with default values of 0.5, reflecting the balance between confidence and error chunk ratio. Mean Confidence is extracted from the model's internal confidence scores, while Error Chunk Ratio represents the proportion of citations that reference incorrect or irrelevant chunks.

\item \textbf{Standard Anomaly} refers to deficiencies in reference answers or evaluation standards. Kamalloo et al~\cite{kamalloo-etal-2023-evaluating} find that discrepancies between system outputs and references often arise from incomplete or incorrect gold standards, with over 50\% of lexical matching failures due to semantically equivalent answers. Researchers advocate for more sophisticated methods to evaluate model generalization and accuracy.

The Standard Anomaly Value combines two components: GPTCheck for uncertainty detection and FactScore for verifying facts against a knowledge base:
\begin{equation}
R_{\text{hall}} = \alpha \cdot \text{GPTCheck}(\text{output}) + \beta \cdot \text{FactScore}(\text{output})
\end{equation}

This metric captures issues like outdated information, erroneous sources, and evaluation set errors. Default values for \(\alpha\) and \(\beta\) are 0.4 and 0.6 respectively, emphasizing factual accuracy over uncertainty detection.
\end{itemize}

\sysname{} supports extensible evaluation \red{metrics}, allowing users to incorporate additional metrics tailored to domain-specific requirements.

\subsection{Data Processing Pipeline}
\label{sec:data_processing}
The data processing pipeline in \sysname{} handles both the initial corpus and runtime queries through a structured workflow. For embedding generation, we utilize OpenAI's text-embedding-3-large model to create high-dimensional vector representations of all text chunks and questions. These embeddings capture semantic relationships between chunks and queries, serving as the foundation for both retrieval and visualization components.

For dimensionality reduction and visualization, we employ openTSNE~\cite{Policar2024}, a modular Python library that implements the t-SNE algorithm. The initial corpus of chunks is projected into a two-dimensional space while preserving local neighborhood relationships:
\begin{equation}
X_{2D} = \text{openTSNE}(X_{emb}, \text{perplexity}, \text{iterations})
\end{equation}

\noindent where $X_{emb}$ represents the high-dimensional embeddings and $X_{2D}$ is the resulting two-dimensional projection. This 2D space is then partitioned into a 200×200 grid, with cell density calculated as:
\begin{equation}
\text{Density}(c_{i,j}) = \frac{|\{x \in X_{2D} | x \text{ falls in cell } c_{i,j}\}|}{|X_{2D}|}
\end{equation}

These density values directly inform the heatmap visualization, with higher density regions rendered with greater intensity.

For thematic analysis and topic extraction within grid cells, we implement a weighted entity-based approach:
\begin{equation}
T(c_i, t) = \sum_{e \in E(c_i)} \text{sim}(e, t) \cdot w(e)
\end{equation}
Where \(T(c_i, t)\) represents the relevance between topic \(t\) and cell \(c_i\). \(E(c_i)\) is the set of named entities extracted from chunks in cell \(c_i\), \(\text{sim}(e, t)\) measures the semantic similarity between entity \(e\) and topic \(t\), and \(w(e)\) is the weighted importance, calculated by TF-IDF value, of entity \(e\) based on its frequency.

For new user queries at runtime, we implement an incremental fitting approach where new embeddings are projected onto the existing t-SNE space. This approach allows new questions to be positioned appropriately relative to the existing knowledge base without recomputing the entire projection. The iteration count (default: 50) can be adjusted by users based on their acceptable latency-quality tradeoff.

% The data flow for each evaluation metric in our system is carefully designed:

% \begin{itemize}[leftmargin=*] 
% \item For \textbf{Retrieval Failure}, the retrieval system produces $C_{ret}$, a set of retrieved chunks for a given query. These chunks are compared against $C_{gold}$, the ground truth chunks annotated by domain experts.

% \item For \textbf{Prompt Fragility}, the system generates $m$ variations of the same query through controlled paraphrasing, feeds each through the retrieval process, and collects the top $n$ chunks for each variation. The resulting chunk sets are then compared pairwise to assess consistency in retrieval outcomes.

% \item For \textbf{Generation Anomaly}, the data flow begins with the generation model producing an answer based on retrieved chunks. The system then extracts confidence signals from the generated text and identifies citation claims. Each citation is verified against the actual content of the referenced chunks, with mismatches counted toward the error chunk ratio.

% \item For \textbf{Standard Anomaly}, the system first processes the generated output through GPTCheck, which identifies uncertain or ambiguous statements. Simultaneously, facts extracted from the output are verified against a curated knowledge base using FactScore.
% \end{itemize}

The data flow for each evaluation metric follows a carefully orchestrated process. For Retrieval Failure assessment, the system compares retrieved chunks against ground truth annotations provided by domain experts. Prompt Fragility evaluation involves generating multiple query variations through controlled paraphrasing, collecting top-ranked chunks for each variation, and performing pairwise comparisons to assess retrieval consistency. Generation Anomaly detection processes model outputs to extract confidence signals and identify citation claims, subsequently verifying each citation against referenced chunk content to calculate error ratios. Standard Anomaly evaluation employs dual verification through GPTCheck for uncertainty detection and FactScore for fact verification against curated knowledge bases.

To provide the most plausible standard chunks in
\red{Chunk-Relink Graph}, we concatenate each question with its corresponding ground truth in the dataset to form a retrieval prompt
that is indexed in the vector database. If no ground truth exists,
users can be guided to define custom standards.
\section{Usage Scenario}  
% Informed by the formative study and design considerations, we propose dynamically visualizing the retrieval and generation dynamics to help users analyze and refine the RAG process. This is achieved through a workflow that identifies retrieval errors, generation inconsistencies, and attribution failures, mapping them to interactive visual components in real time (see~\cref{fig:system_main}). These components not only illustrate the static aspects of retrieval and generation but also display the evolving retrieval rankings, chunk attributions, and answer compositions throughout the pipeline. Moreover, they provide users with rich interaction possibilities, allowing them to refine retrieval strategies and contextual augmentation without modifying the underlying model.
% We instantiate this idea based on \sysname{} with a prototype system, which enables users to proactively investigate and optimize retrieval-generation interactions.
This section walks through \sysname{} using hypothetical use cases, demonstrating its capacity to enhance transparency and control in RAG-based systems.

\begin{figure}[!b]
	\centering	
	\includegraphics[width=1.0\columnwidth]{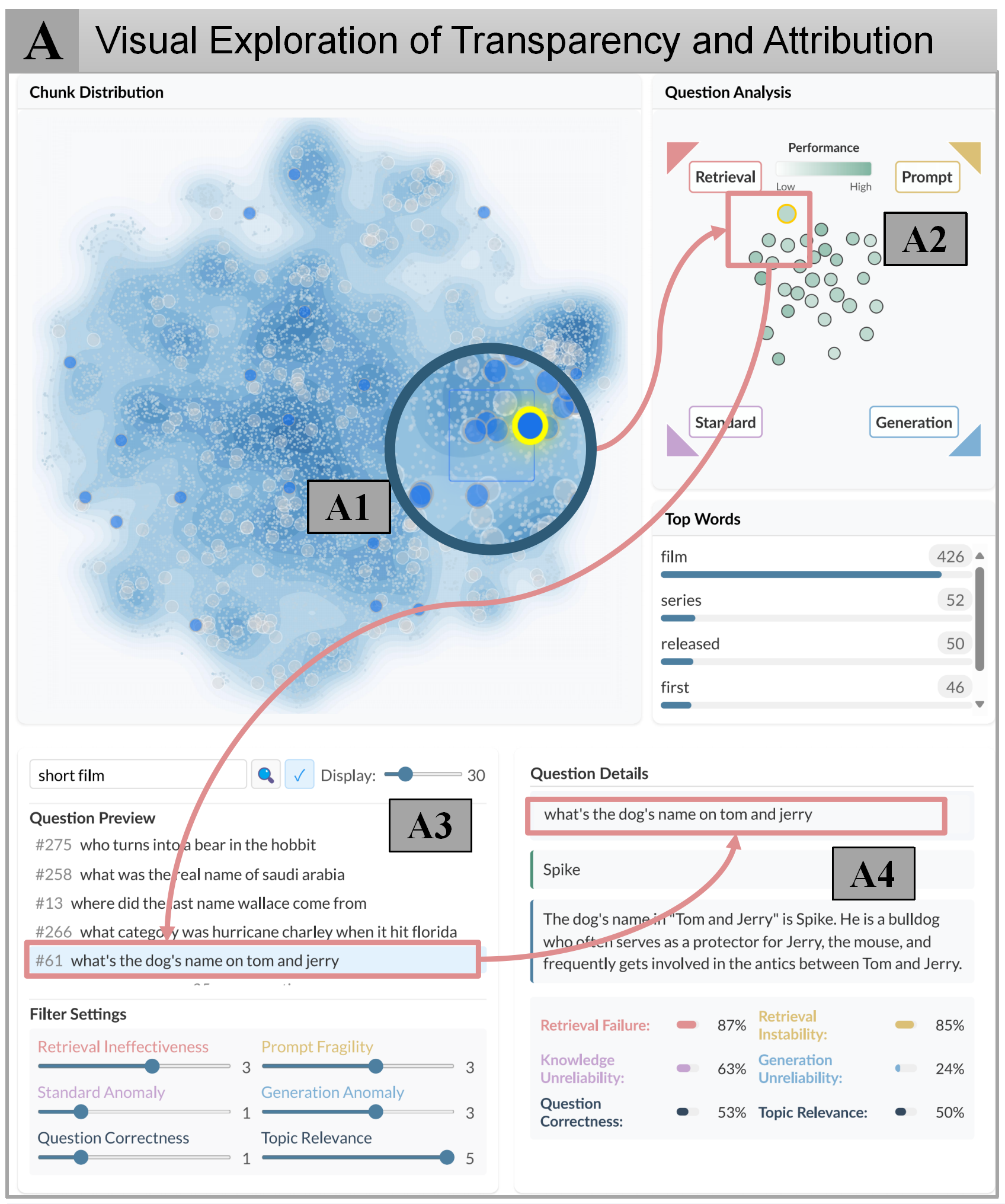}
	\caption{In Heatmap (A1), Force-directed Graph (A2), Question Search (A3) and Question Details (A4), the same question is represented using different visual encodings.}
    \label{fig:moduleA}
\end{figure}

\begin{figure}[!b]
	\centering	
	\includegraphics[width=1.0\columnwidth]{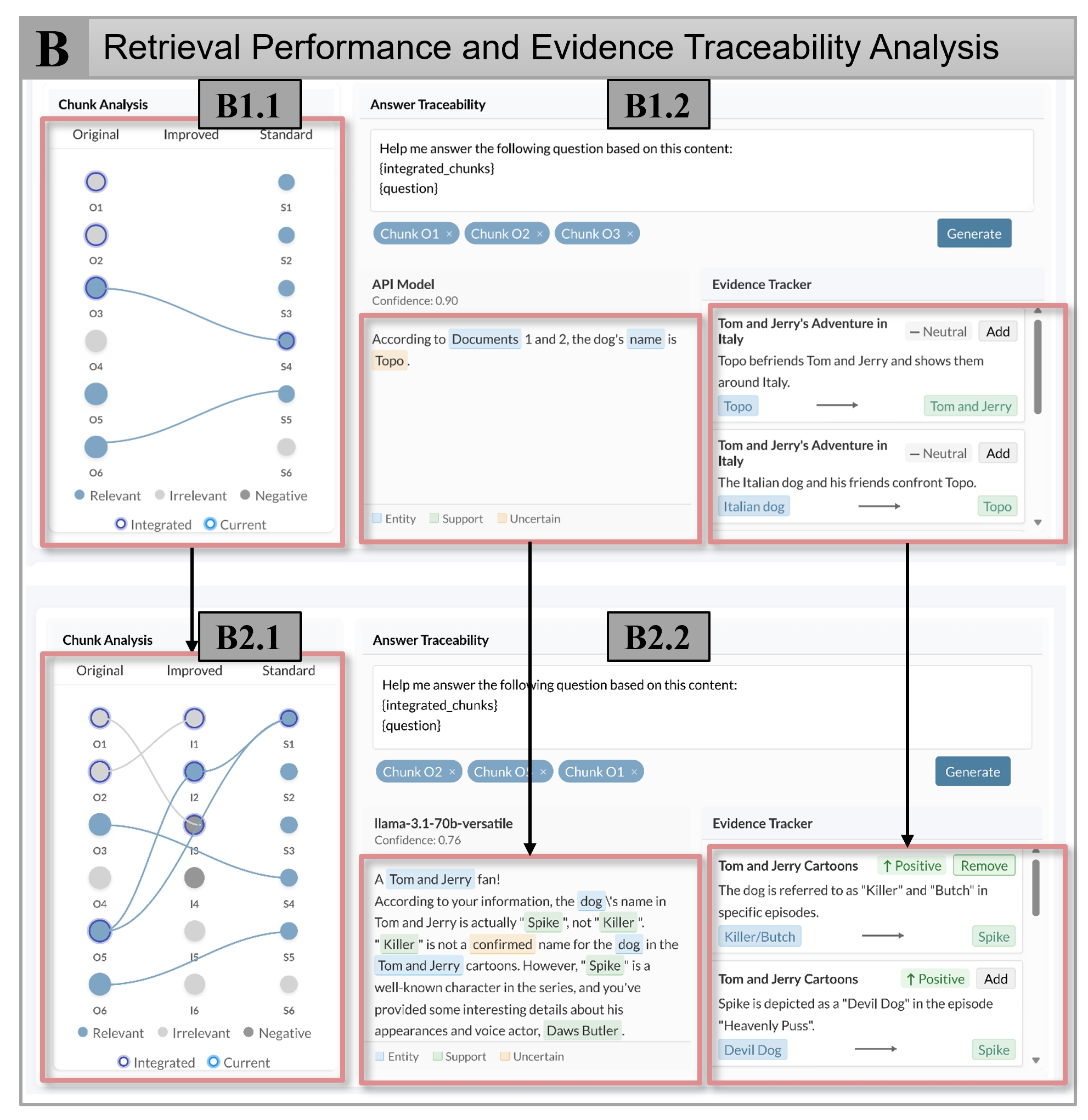}
	\caption{Changes in the \red{Chunk-Relink Graph}, corresponding evidence chains, and model-generated answers before and after user tuning.}
    \label{fig:moduleB}
\end{figure}

\begin{figure}[!t]
	\centering	
	\includegraphics[width=1.0\columnwidth]{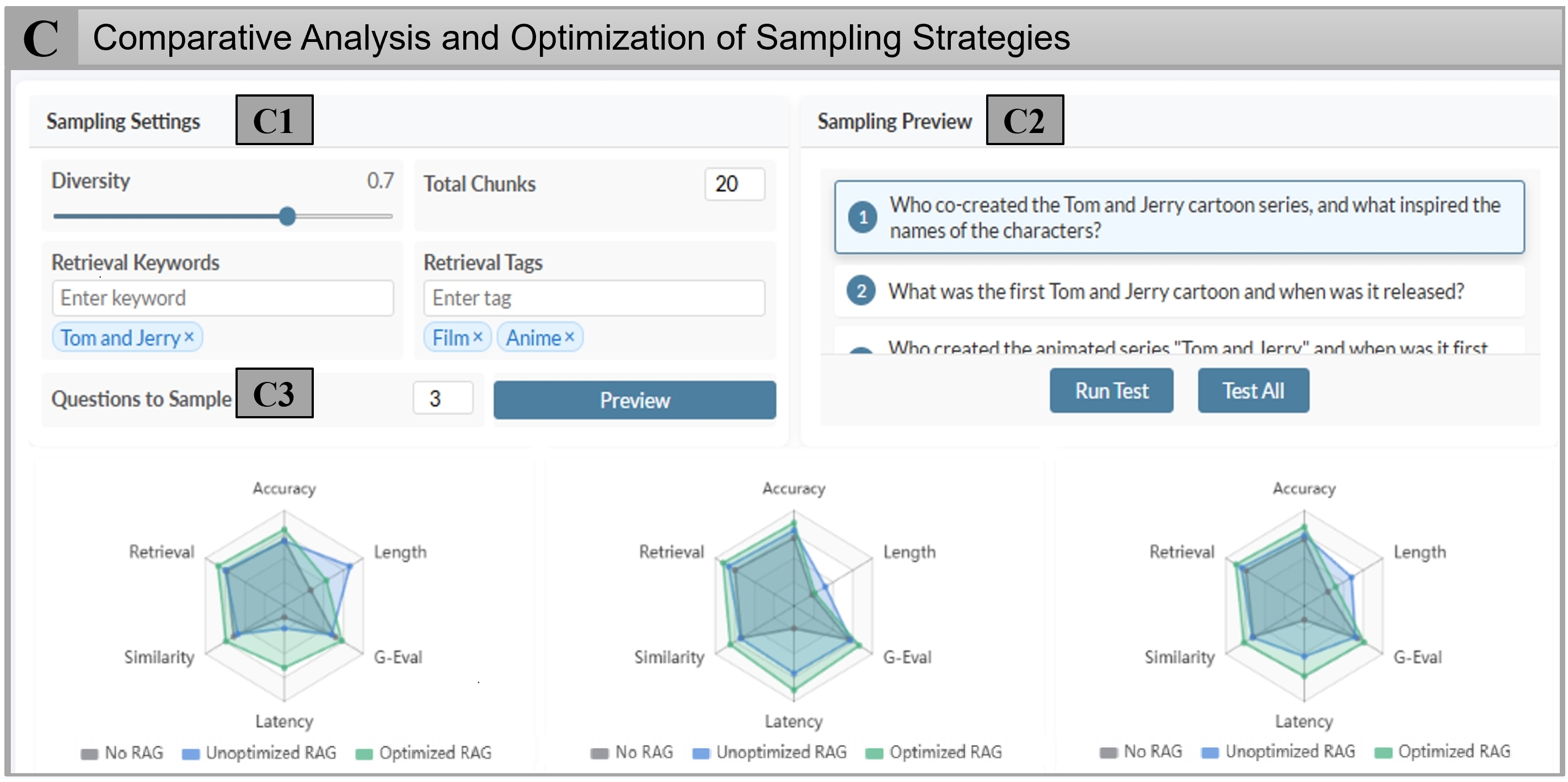}
	\caption{The new questions generated by users regarding the ``Tom and Jerry'' question, as well as their performance variations before and after \red{optimization}.}
    \label{fig:moduleC}
\end{figure}

\vspace{1.2mm}\noindent\textit{\textbf{Scenario Setting.}} 
Alex, a researcher exploring the performance of RAG systems in handling factual questions, turns to \sysname{} to investigate retrieval inconsistencies. He is particularly interested in identifying issues where entity fragmentation leads to inaccurate or incomplete responses. To explore this, Alex examines the question: “What’s the dog’s name in Tom and Jerry?” - a seemingly simple fact-based inquiry that poses challenges for retrieval and generation consistency.

\vspace{1.2mm}\noindent\textit{\textbf{Visual Exploration of Transparency and Attribution.}}
Upon launching \sysname{}, Alex engages in a structured exploration of transparency and attribution using the visual diagnostics interface in ~\cref{fig:moduleA} (A). He begins with the chunk-level Heatmap view (A1), which presents an overview of retrieval activity across documents. By entering ``short film'' in the Question Search (A3) to perform an embedding-level search, Alex observes a densely populated cluster of retrieved chunks, suggesting concentrated retrieval behavior in a specific domain. To investigate further, he transitions to the Force-directed Graph view (A2), which encodes similarity relationships among queries. Here, he identifies anomalous patterns in a subset of queries that diverge from canonical retrieval paths. Notably, the query ``What’s the dog’s name in Tom and Jerry?'' resides at the intersection of two distinct failure clusters (\red{\textbf{Retrieval Failure} and \textbf{Prompt Fragility}}), suggesting it embodies characteristics of multiple retrieval error modes. Continuing the analysis, Alex uses the entity-level diagnostic view (A4) and uncovers a recurring issue of entity fragmentation across semantically related queries. Together, these coordinated views (A1-A4) offer a comprehensive understanding of retrieval behaviors and failure patterns. As illustrated in ~\cref{fig:moduleA}, this workflow demonstrates how \sysname{} enables users to systematically trace and interpret retrieval dynamics through integrated visual analysis.

\vspace{1.2mm}\noindent\textit{\textbf{Retrieval Performance and Evidence Traceability Analysis.}}
Intrigued, Alex moves to the Retrieval Performance and Evidence Traceability Analysis in~\cref{fig:moduleB} (B) to inspect the document ranking visualization. After \red{automatic cross-component synchronization}, he quickly noticed a blue chunk in the third row of the ``standard'' field, suggesting a close match to his desired answer. Using \sysname{}'s \red{Chunk-Relink Graph},~\cref{fig:moduleB} (B1.1), Alex observes that the system's retrieval step did not prioritize this chunk effectively. Correlating this with the previously identified retrieval anomalies, he cross-references the Force-directed Graph and confirms that the retrieval ranking inconsistencies align with the observed entity fragmentation patterns. The Evidence Tracker and the model's answer reveal that information related to ``Spike'' is scattered across multiple retrieved documents, with its information density diluted by irrelevant documents,~\cref{fig:moduleB} (B1.2).

Alex narrows the broad issue of indicator anomalies down to retrieval anomalies and overly scattered retrieval content, possibly due to the lack of relevance labeling in the data source. So, he tries to recall algorithms that address such fine-grained problems. He applies the HyDE~\cite{gao-etal-2023-precise} (Hypothetical Document Embeddings) method, which generates a hypothetical answer document based on the question and uses its embedding for retrieval. As shown in~\cref{fig:moduleB} (B2.1), switching to the multi-round retrieval visualization, he tracks how chunk rankings evolve. He observes that after applying HyDE, the system gradually prioritizes previously overlooked “Spike” chunk, which actually contained complete information on the topic he wanted,~\cref{fig:moduleB} (B2.2), rather than fragmented information.

\vspace{1.2mm}\noindent\textit{\textbf{Comparative Analysis and Optimization of RAG Sampling Strategies.}}
To evaluate the effectiveness of this optimization, Alex transitions to the Comparative Analysis and RAG Sampling Optimization in~\cref{fig:moduleC} (C). Using the Sampling Settings panel,~\cref{fig:moduleC} (C1), he configures parameters to identify similar entity-centric questions with potential retrieval inconsistencies. He selects queries related to entertainment media and character identification, which might suffer from similar fragmentation challenges~\cref{fig:moduleC} (C2).

The system presents these sampled questions alongside a Radar Chart visualization,~\cref{fig:moduleC} (C3), enabling Alex to compare performance metrics before and after optimization. The visualization highlights a consistent improvement in factual correctness and BLEU scores, particularly for queries that previously exhibited entity fragmentation. Further comparative analysis using multiple RAG strategies confirms that improved chunk consolidation and advanced retrieval techniques significantly enhance response accuracy, as illustrated in ~\cref{fig:moduleC}.

\vspace{1.2mm}\noindent\textit{\textbf{Domain Expert Workflow for RAG Optimization.}}
To illustrate the system's utility for users without deep RAG algorithm knowledge, consider Robin, a medical researcher unfamiliar with RAG optimization intricacies. Robin uses \sysname{} to investigate potential causes for a patient's symptoms by identifying the question: ``A 45-year-old male patient complains of intermittent chest pain, with the pain concentrated behind the sternum. What could be the possible cause?''.

\begin{figure}[!t]
	\centering
	\includegraphics[width=1.0\columnwidth]{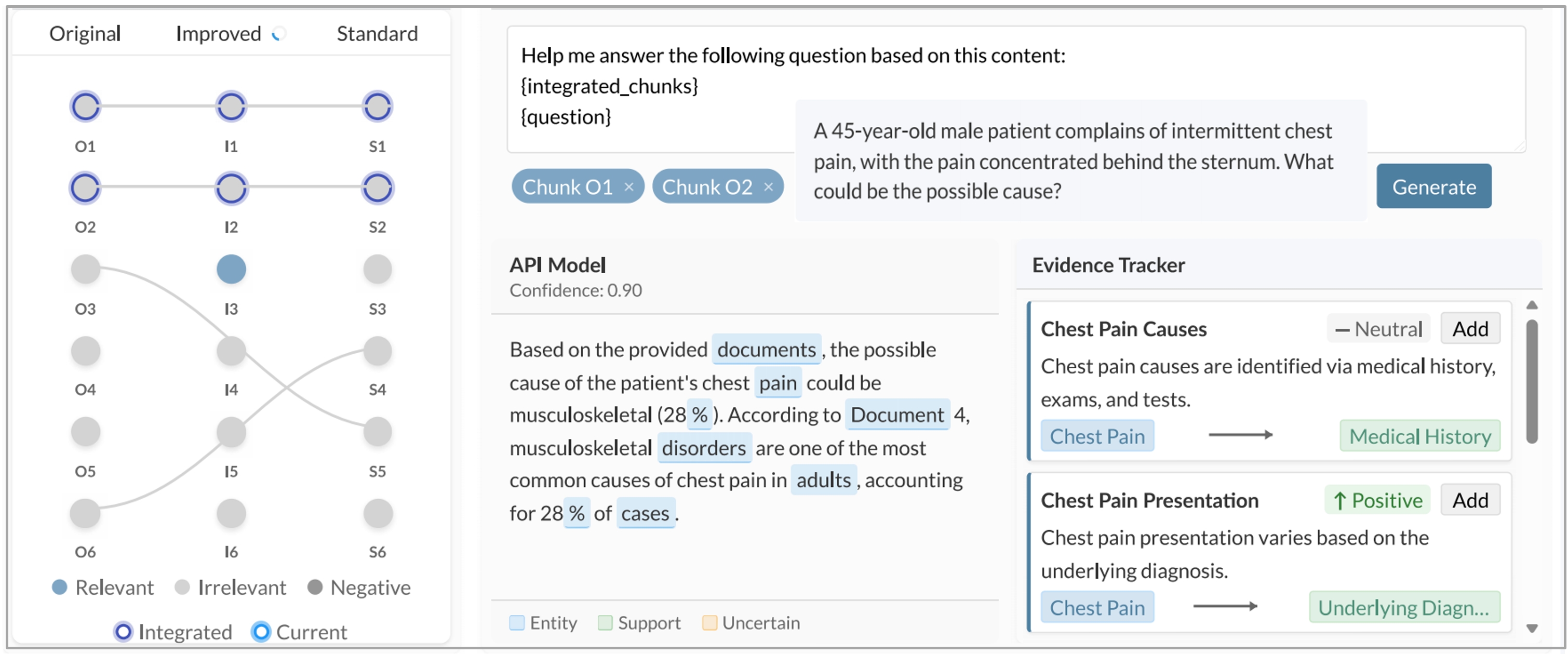}
	\caption{Robin's case: Initial retrieval results (left) versus results after prompt refinement (right), showing improved but still suboptimal answer generation.}
    \label{fig:moduleD}
\end{figure}

Transitioning to~\cref{fig:moduleD}, Robin examines the initial retrieval results in the \red{Chunk-Relink Graph},~\cref{fig:moduleD} (B1). He finds the top retrieved chunks unsatisfactory, offering only generic information about ``chest pain'' without specific diagnostic pointers. Guided by the system, Robin refines the retrieval prompt in~\cref{fig:moduleD} (B2), incorporating his domain expertise. He adds phrases like ``possible differential diagnoses for substernal chest pain'' and ``cardiac conditions presenting with atypical symptoms''. The system then retrieves slightly more relevant information, and the generated answer improves, but still lacks the required clinical depth and precision.
% , ~\cref{fig:moduleD}.

Recognizing that the general RAG data source might lack specialized medical knowledge, Robin decides to augment the knowledge base. He incorporates a more domain-specific dataset, such as the PubMed Central Open Access dataset\footnote{https://pmc.ncbi.nlm.nih.gov/tools/openftlist/}. After integrating this new data source, \sysname{}'s retrieval mechanism accesses more precise information. The \red{Chunk-Relink Graph} now highlights documents discussing conditions like microvascular angina and anxiety-related chest pain, leading to a significantly more accurate and clinically relevant answer.

\begin{figure}[!t]
	\centering
	\includegraphics[width=0.45\columnwidth]{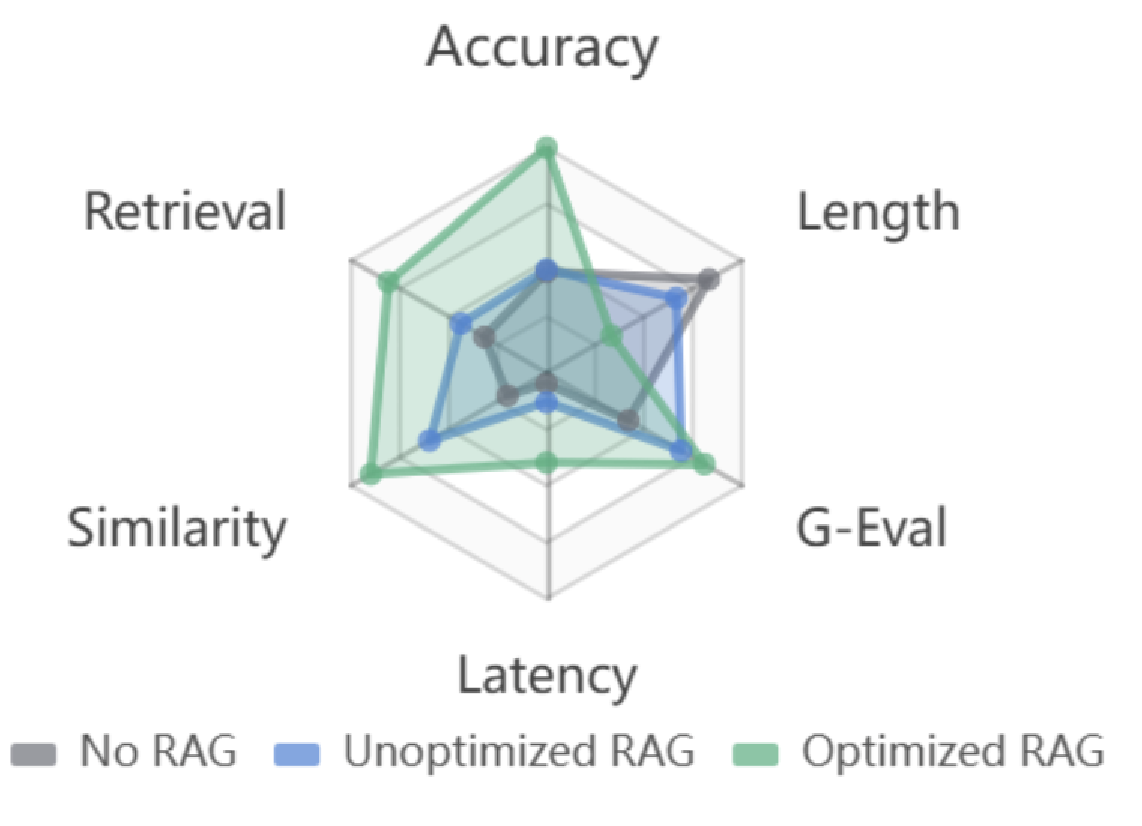}
	\caption{Robin uses the Radar Chart to validate improved accuracy and relevance metrics.}
    \label{fig:robin}
\end{figure}

Similar to Alex's approach, Robin transitions to the Comparative Analysis and Optimization component to evaluate his optimizations. He configures the Sampling Settings panel to test similar medical diagnostic questions. The Radar Chart visualization reveals substantial improvements in clinical accuracy and relevance scores across the sampled questions, ~\cref{fig:robin}. Encouraged by these results, Robin can iteratively return to the two previous components to further refine his approach, creating a continuous improvement cycle.
%This iterative workflow demonstrates how \sysname{} empowers domain experts to leverage their specialized knowledge to enhance RAG performance through prompt refinement and data source integration, even without expertise in the underlying algorithms.

Through \sysname{}, Alex and Robin effectively diagnose and mitigate entity fragmentation and document missing issues in RAG-based queries, leading to more reliable and precise information retrieval. This workflow exemplifies how interactive visual analysis can empower users to refine retrieval strategies and enhance system transparency.
\section{User Study}

To understand how \sysname{} supports users in diagnosing and refining retrieval-generation dynamics in RAG systems, we conducted a \red{user study} where participants used \sysname{} to analyze and debug RAG systems. Our goal was to explore how the interactive visualization tools in \sysname{} assist users in identifying issues, interpreting model behavior, and iterating retrieval strategies.

In this study, we aimed to answer the following research questions: 
\begin{itemize}[leftmargin=*] 
    \item \textbf{RQ1.} How does \sysname{} help users identify attribution failures, entity fragmentation, or retrieval inconsistencies in RAG systems?
    \item \textbf{RQ2.} In what ways does \sysname{} influence users' strategies for debugging and optimizing RAG workflows, particularly in terms of chunk selection, retrieval refinement, and prompt adaptation? 
    \item \textbf{RQ3.} How do users interpret and make use of the visual explanations provided by \sysname{}, and to what extent do they trust the insights generated across different views (e.g., retrieval ranking, chunk relinking, generation anomaly graphs)? 
\end{itemize}

\subsection{Study Design}
\subsubsection{Participants.}
We recruited 11 participants (8 males, 3 females) through surveys and referrals. All participants had prior experience working with LLMs, particularly in contexts involving RAG. Four participants (P1-P4) are RAG researchers familiar with prompt engineering and information retrieval, while seven of them (P5-P11) are experts in various research fields. Participants were compensated with local currency equivalent to \$15 for their participation.

\subsubsection{Procedure.}
Participants first signed an informed consent form before beginning the study. Each session began with a 5-minute introduction to the \sysname{} \red{system}, providing an overview of its key features and functionality. Participants were then asked to select three questions of interest from the Natural Questions (NQ)\footnote{The Natural Questions (NQ) corpus is a real-user open-domain QA dataset where systems must comprehend entire Wikipedia articles to answer questions, designed for realistic and challenging evaluation.} ~\cite{nq} dataset to evaluate the performance of the Llama3-70B model with a RAG system. \red{The study utilized a randomly sampled subset of 300 questions from NQ test set}. \red{The RAG system was configured with a Wikipedia-based knowledge base containing approximately 20 million semantically segmented chunks, ensuring comprehensive coverage of factual information across multiple domains.} 

For each selected question, participants were instructed to assess the model's performance, identify potential issues in cases where the model performed poorly, and propose insights regarding possible strategies to enhance the RAG workflow. After evaluating the questions, participants were asked to improve the model's generation quality by adjusting retrieval parameters, refining generation prompts, or incorporating external or custom data sources. They then used \sysname{}'s iterative system evaluation to assess the actual performance improvements resulting from these modifications. When participants encountered retrieval or generation failures, they were encouraged to analyze the underlying causes and suggest actionable refinements to improve system performance.

After completing the assigned tasks, participants were given time to freely explore the \sysname{} tools, investigating additional questions or features according to their interests. The sessions concluded with a semi-structured interview where participants shared their comprehensive impressions of the tool and evaluated its effectiveness in diagnosing and refining RAG systems. Study sessions ranged from 45 to 90 minutes in duration, depending on the depth of participants' exploration and discussion.

\subsubsection{Measures.}
For qualitative data, we transcribed participants' responses from the semi-structured interviews and applied thematic analysis to extract key themes related to their reasoning strategies, trust in \sysname{}, and refinement approaches. Two researchers independently coded the transcripts and resolved discrepancies through discussion.

\begin{figure*}[]
    \centering
    \includegraphics[width=1.00\textwidth]{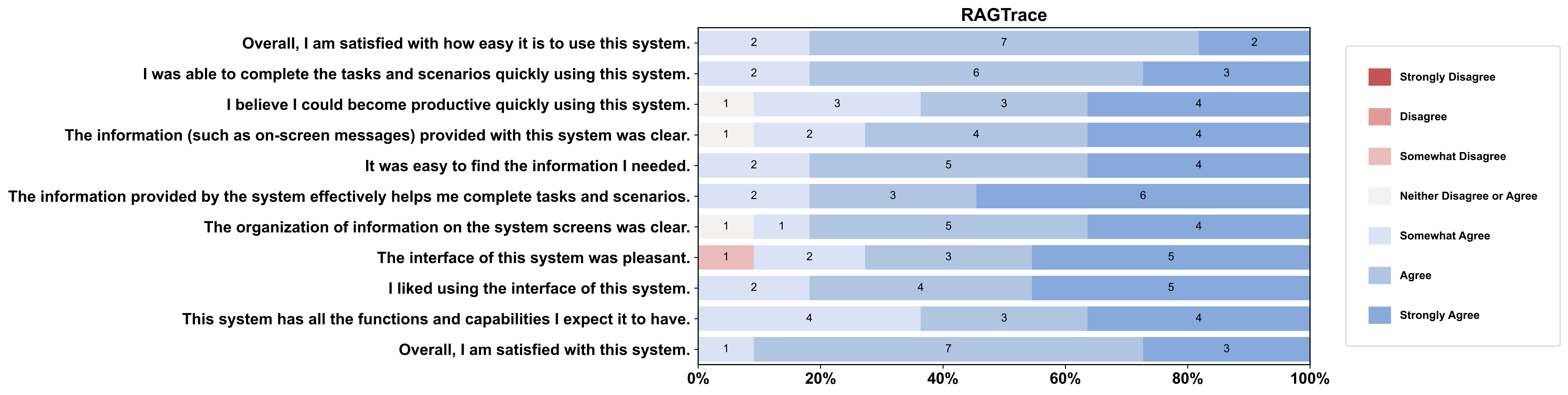}
    \caption{Participant Responses to Post-Study System Usability Questionnaire (PSSUQ) }
    \label{fig:PSSUQ}
\end{figure*}

For quantitative data, we used six items from the NASA-TLX questionnaire (including Physical Demand). Notably, we reversed the scale for the "Performance" dimension (related to "How successful were you in accomplishing what you were asked to do?") to align with users' intuitive expectations, so that higher values indicate greater perceived success, enhancing interpretability of the results. Additionally, we employed the Post-Study System Usability Questionnaire (PSSUQ) to evaluate participants' subjective satisfaction with \sysname{}'s usability, interface design, and overall functionality. The PSSUQ provided valuable insights into users' perceived effectiveness and satisfaction with the system across multiple dimensions. Full survey items are listed in ~\cref{fig:PSSUQ}. Likert-scale responses were analyzed using the Wilcoxon signed-rank test due to the ordinal nature of the data.

We did not conduct external evaluations of participants' refinements, as debugging RAG workflows is highly contextual. Participants set their own debugging goals, and external criteria could lead to misaligned assessments. Their explanations also involved subjective interpretations of relevance and hallucination, making external judgment unreliable.

\subsection{Results}

\begin{figure*}[]
    \centering
    \includegraphics[width=1.00\textwidth]{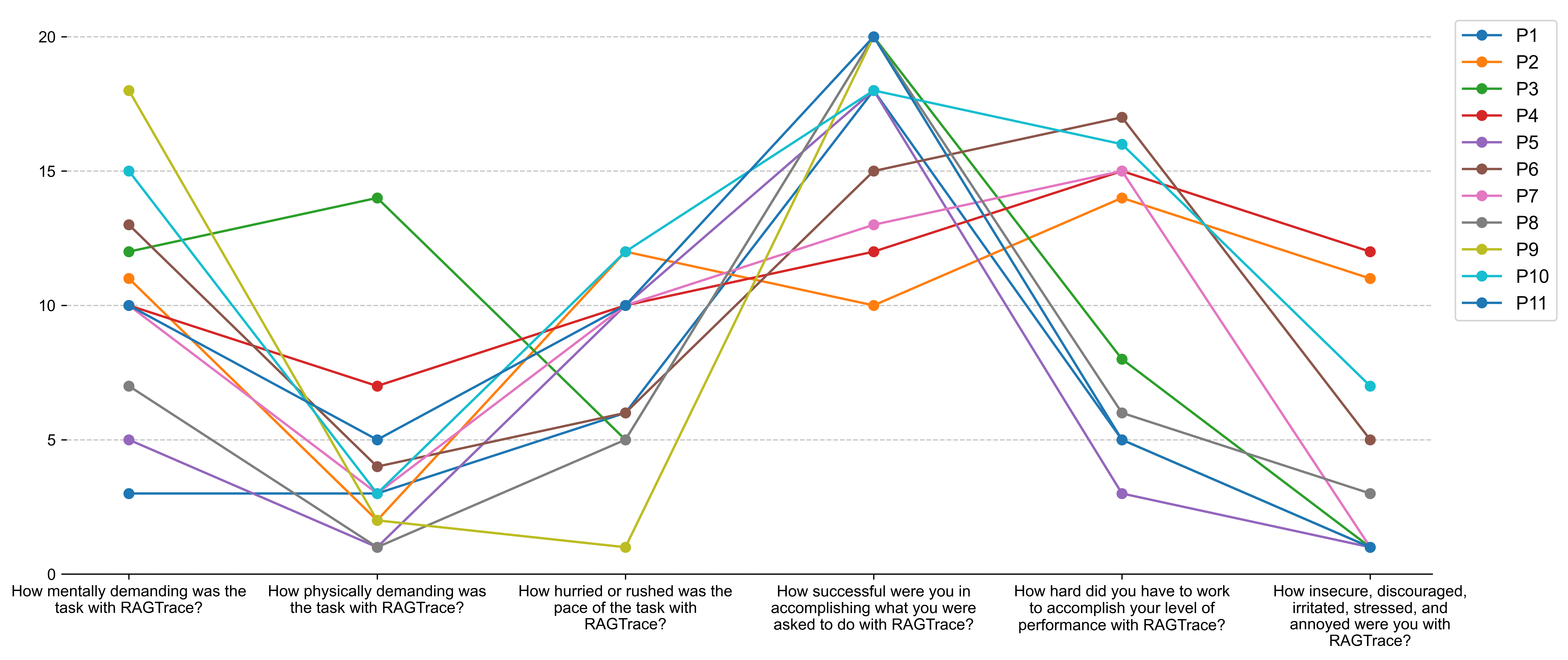}
    \caption{Participant Responses to the NASA-TLX Cognitive Load Questionnaire}
    \label{fig:NASA}
\end{figure*}

Our quantitative analysis revealed high user satisfaction with \sysname{}'s usability and effectiveness in supporting RAG system diagnosis and refinement. The PSSUQ results indicated strong overall satisfaction (mean = 6.18, SD = 0.60 on a 7-point scale), addressing \textbf{RQ3} by confirming users' positive reception of the system's visual explanations. Participants particularly valued the system's ability to effectively help them complete tasks (mean = 6.36, SD = 0.81) and its pleasant interface (mean = 6.27, SD = 0.79). The NASA-TLX results showed moderate mental demand (mean = 10.36, SD = 4.30 on a 20-point scale), low physical demand (mean = 4.09, SD = 3.73), and notably high self-reported task success (mean = 16.73, SD = 3.64), the latter supporting \textbf{RQ2} by demonstrating that users could effectively implement optimization strategies using the system. Participants also reported low stress levels (mean = 4.00, SD = 4.22) despite engaging with complex analytical tasks, suggesting that \sysname{} successfully reduced cognitive burden when analyzing RAG workflows. The combination of high usability ratings with moderate workload metrics indicates that \sysname{} effectively balances analytical power with accessibility, allowing both RAG experts and domain specialists to effectively leverage the system's capabilities.

Based on the semi-structured interviews, the feedback revealed several key benefits and areas for improvement regarding \sysname{}. Participants consistently reported that the system enhanced their ability to understand, debug, and refine RAG workflows.

\subsubsection{Enhanced Understanding and Debugging Capabilities.}
In response to \textbf{RQ1}, participants consistently highlighted \sysname{}'s effectiveness in demystifying the complex retrieval-generation process. In particular, the evidence chain visualization was frequently praised for its utility in tracing the origin of generated answers and verifying their correctness. As P7 articulated, it helped determine ``if the answer is genuinely correct or just superficially so,'' providing crucial insights into the ``source of the retrieved answer and its validity.'' P5 also found this feature ``very practical for clearly understanding the pipeline.'' The system was noted for surfacing issues that might otherwise go unnoticed (P4). For instance, P9 utilized the evidence traceability analysis component to understand how query modifications (like adding quotes or themes) influenced retrieval outcomes, finding it particularly valuable when ``embedding models are unstable,'' enabling them to ``retrieve useful knowledge through experimentation.'' P9 further appreciated the system's ability to ``supplement missing information, like the subject,'' in fragmented text outputs from the LLM. This deeper understanding directly facilitated more effective debugging strategies. P10 mentioned using the tool to diagnose ``whether the knowledge base was lacking content'' or assess ``where RAG is useful,'' informing subsequent prompt improvements. Participants felt the system provided a much clearer grasp of the overall RAG workflow (P3, P9), with P9 contrasting it favorably against previous experiences of manually feeding knowledge to models.

\vspace{1.2mm}\noindent\textbf{System Intuitiveness and Responsiveness.}
Participants commended the system's intuitive design and responsive interface. P5 described the system as having ``very good responsiveness, with operations feeling smooth and natural,'' noting that ``the interface design is clear with well-defined functional modules, making the entire system easy to comprehend.'' P4 emphasized how the structured layout effectively linked inputs to analytical outputs, significantly reducing debugging time. The system was generally characterized as ``clear and responsive,'' with P5 adding that it became ``quite clear after familiarization.'' The color design was particularly praised for its clarity and visual satisfaction. P3 appreciated how the system provided a clearer understanding of ``the overall workflow,'' while P6 highlighted the logical coherence of the system's process flow. P1 valued how the system ``clearly separates various functional modules, allowing users to utilize them as needed.'' Overall, participants found that the visualization components significantly enhanced their understanding of the complex RAG processes.

\subsubsection{Improved Efficiency and Workflow Integration.}
In response to \textbf{RQ2}, participants commonly reported substantial improvements in their debugging strategies and optimization workflows, particularly in terms of chunk selection, retrieval refinement, and prompt adaptation, when using \sysname{}.
P7 estimated an efficiency improvement of retrieving document ``at least 80\%'' compared to their previous methods without the tool. P4 emphasized that the structured layout, linking inputs to analytical outputs, ``effectively reduces time'' spent debugging. P9 reflected that the tool significantly enhanced their understanding of RAG, appreciating how it enables leveraging external knowledge to ``improve the model's results when its own knowledge is limited.'' Furthermore, RAGTrace was seen as a valuable asset for demonstrating RAG system behavior and effectiveness (P2). P6 noted that the comprehensive functionality covering ``retrieval, fine-tuning, and analysis'' made the system applicable across multiple scenarios, from academic research to industrial optimization.

Opinions varied regarding the tool's integration into existing workflows. While industry practitioners (P6, P2) were enthusiastic about its potential for immediate integration into production environments, academic researchers (P3, P9) expressed more measured views about its applicability across different research contexts. P3 commented that ``while extremely valuable for targeted diagnostics, the system might require customization for novel RAG architectures that deviate from standard retrieval patterns.'' This distinction highlights the different requirements between standardized production environments and more experimental research settings.

\subsubsection{User Exploration Strategies and Insights.}
In response to \textbf{RQ3}, participants' interaction feedback revealed distinct exploration and sensemaking patterns, with users leveraging different components of \sysname{} to interpret retrieval-generation dynamics and develop trust in the system's multi-faceted visual explanations.

\vspace{1.2mm}\noindent\textbf{Interactive Question Discovery and Location.} 
Participants valued \sysname{}'s capabilities for question exploration and rapid location. The force-directed graph was described by P4 as effectively facilitating \textit{``navigation and locating desired questions,''} while P2 appreciated how it helped \textit{``position questions to find the most relevant ones.''} P6 considered the search questions feature a \textit{``core function''}, and P7 emphasized the system's ability to \textit{``generate new questions and add custom queries beyond the static dataset,''} enabling \textit{``dynamic assessment of system performance.''} When combined with the heatmap, these tools provided powerful problem identification capabilities, with P4 noting that this combination could effectively reveal \textit{``issues that might otherwise go unnoticed.''} P5 stated that the heatmap provided a \textit{``clear cognitive map''} for identifying problematic terms to avoid, particularly how it visualized \textit{``white spots that could be used to avoid certain specific keywords.''} These exploration patterns demonstrate \sysname{}'s multi-layered support for question discovery and location.

\vspace{1.2mm}\noindent\textbf{Query Reformulation and Retrieval Strategy Experimentation.} 
Multiple participants engaged in systematic experimentation with query formulations to understand retrieval behavior. P9 discovered meaningful patterns through the system: \textit{``After using this tool, I found that adding quotes around key terms in prompts produces different results, and introducing thematic elements seems to change outcomes as well.''} Through knowledge base evaluation, P9 also revealed insights about embedding model instability: \textit{``I've realized that embedding models can be unstable—trying different chunks a few times might eventually find the correct answer.''} P10 and P1 employed the system to identify and remove redundant or misleading information from the knowledge base. P10 noted that \textit{``cleaning up noisy documents improved retrieval precision more than adding additional context in several test cases,''} representing an alternative optimization strategy that some participants discovered through system exploration. However, P3 disagreed with this approach, arguing that \textit{``maintaining comprehensive coverage is often more important than removing noise, since the ranking algorithm should handle relevance sorting appropriately.''} This difference in perspective reflected participants' diverse backgrounds and the varied requirements of their use cases, while demonstrating the system's flexibility in supporting different retrieval strategy experiments.

\vspace{1.2mm}\noindent\textbf{Attribution and Evidence Chain Analysis.} 
Participants found the system effective in revealing answer sources and verifying correctness. P2 particularly valued the \red{Chunk-Relink Graph's} comparison between standard and retrieved chunks, noting it was \textit{``most useful''} because it visually revealed \textit{``which standard answers were not utilized,''} thus identifying \textit{``what information was lost''} and providing clear \textit{``directions for improvement.''} P1 also praised the \red{Chunk-Relink Graph} for \textit{``clarifying several types of problems in one view,''} while P6 highlighted that the \red{Chunk-Relink Graph} helped \textit{``quickly locate specific chunks when diagnosing retrieval-based errors.''} P7 found the evidence chain visualization effective for tracing the origin of generated answers and verifying their correctness, helping determine \textit{``if the answer is genuinely correct or just superficially so,''} providing crucial insights into the \textit{``source of the retrieved answer and its validity.''} P5 also found this feature \textit{``very practical for clearly understanding the pipeline.''} These responses demonstrate \sysname{}'s value in providing detailed diagnostic capabilities and transparency.

\vspace{1.2mm}\noindent\textbf{Iterative Performance Analysis and Optimization.} 
Participants evaluated and optimized system performance through complementary visualization components. The Radar Chart was utilized by P10 to effectively \textit{``compare performance before and after improving prompts,''} while P8 appreciated its ability to facilitate \textit{``associations between related questions,''} and P11 observed that \textit{``when indicator values change, the Radar Chart might provide hints about problem locations.''} P11 particularly appreciated how the system provided \textit{``direct descriptive terminology that predefined good performance,''} reducing cognitive load by \textit{``helping users determine what constitutes good performance without extensive analysis,''} while still allowing them to \textit{``investigate how conclusions were reached.''} P3 valued the ability to quickly grasp \textit{``the overall workflow more clearly,''} demonstrating \sysname{}'s value in providing high-level performance perspectives while supporting iterative system improvements. P9 contrasted this exploration pattern favorably against previous experiences: \textit{``Before using this, I never worked with RAG systems—I always manually fed knowledge to models. Now I understand how to leverage systems to process knowledge that may not be initially verifiable.''} This exploration pattern shows \sysname{}'s capacity to expand users' conceptual understanding of RAG while supporting practical comparative analysis and iterative optimization.

\subsubsection{System Limitations.}
Participants provided constructive feedback for future enhancements. Increased interactivity was a common theme, with suggestions to add previews of text content on hover/click within visualizations (P6) and functionality to highlight corresponding elements (e.g., selected nodes, chunks) across different views (P2). Customization options were also desired, including the ability to define ``custom metrics'' (P3), apply ``directional filtering'' to searches or visualizations (P1), and modify how metrics are displayed on the Radar Chart (P2). P11 suggested incorporating ``pre-defined interpretations'' or hints, possibly on the Radar Chart, to ``reduce cognitive load'' by offering initial assessments of performance changes, guiding further investigation. Lastly, enhancing text exploration by highlighting searched terms within retrieved chunks (P2) and making chunks easily clickable to view original context (P6) were also proposed.

Participants' improvement suggestions sometimes reflected competing priorities. For instance, while some participants (P11, P3) advocated for more automated guidance and interpretations to reduce cognitive load, others (P10, P7) emphasized the importance of maintaining user control and avoiding excessive automation that might obscure important nuances. P10 specifically cautioned against ``over-abstracting the underlying retrieval mechanics,'' noting that ``sometimes the `messiness' of the raw retrieval results contains important diagnostic information.'' These contrasting perspectives highlight the challenge of balancing automation and transparency in analytical systems designed for diverse user groups with varying levels of technical expertise.

\section{Discussion and Conclusion}
% We discuss how \sysname{} helps users analyze retrieval-generation dynamics, balance interpretability with system complexity, and leverage domain expertise to refine retrieval and generation strategies.

In this paper, we present \sysname{}, an interactive evaluation \red{system} to support diagnosis and optimization of RAG workflows. Based on a formative study (N=12) with RAG users, we distill key practices, challenges, and expectations in understanding retrieval relevance, tracing knowledge propagation, and resolving generation inconsistencies. Building on these insights, we develop a diagnostic methodology and an interactive analysis system that enable users to assess retrieval quality, identify generation errors, and refine retrieval strategies for improved RAG performance. A user study (N=11) and expert interviews demonstrated that \sysname{} could effectively facilitate troubleshooting, enhance users' understanding of retrieval-generation interactions, and foster more reliable and controllable RAG workflows.

\subsection{Comparison with Conventional Evaluation Approaches}
Traditional evaluation practices for RAG systems typically focus on isolated metrics for retrieval precision or generation quality~\cite{Zamani2022Retrieval, yu2024evaluation}. In contrast, \sysname{} provides an integrated analytical environment that enables users to explore the full spectrum of retrieval-generation relationships. By visualizing high-level performance metrics alongside fine-grained analyses of retrieval relevance, generation fidelity, and cross-component interactions, our \red{system} facilitates a deeper understanding of the influence that external knowledge sources exert on generated outputs. This multi-level approach addresses the research gaps identified in previous studies and extends the capabilities of conventional toolkits by emphasizing transparency in internal knowledge integration.

\subsection{Design Implications}
\subsubsection{Understanding RAG Interactions Via Transparent Visualization.}
\sysname{} provides real-time visualization of the retrieval-generation dynamics, enabling users to clearly trace how retrieved chunks influence the generated content across iterations. This design echoes the ``visible hands'' approach proposed in prior work (e.g., WaitGPT~\cite{Xie2024uist}), where abstracted operations and dynamic visual feedback help externalize the behavior of LLM agents for better user understanding. In line with research advocating for interpretable and user-steerable AI systems, \sysname{} demonstrates how visualizing intermediate retrieval results and generation paths can foster user trust and facilitate more informed interactions.

\subsubsection{Scrollytelling for Chunk Iteration.}
\red{Building on scrollytelling techniques, which dynamically reveal content through user scrolling to enhance engagement with LLM-generated outputs ~\cite{Xie2024uist}, \sysname{} adapts this approach to visually trace the iterative refinement of retrieved chunks during RAG processes.} This progressive and contextual representation allows users to better understand how the system refines its responses over time, aligning with design guidelines for incremental information disclosure in interactive systems~\cite{osti_1644018}. Such design implications can inspire future RAG interfaces to support transparent and user-friendly exploration of retrieval and generation dynamics.

\subsection{Scalability and Adaptability} 
\sysname{} is designed to operate effectively across a range of system scales—from standard laboratory experiments to real-world deployments with large-scale document repositories. While current implementations manage typical retrieval volumes and generation tasks efficiently, challenges such as increasing log data and the risk of visual clutter in dense scatterplot visualizations remain. Future enhancements will explore advanced sampling and visualization simplification techniques to ensure that our \red{system} remains robust and responsive even under high-demand conditions.

\subsection{Generalizability and Extensibility} 
The data abstraction and multi-level analytical \red{system} underlying \REVISE{\sysname{} is} inherently adaptable. Although we have demonstrated its effectiveness using several real-world RAG applications, the \red{system's} modular design allows it to be extended to various domains and customized for different retrieval and generation strategies. For example, while our current implementation effectively tracks token-level interactions and generation fidelity, adapting the \red{system} to support specialized retrieval strategies (e.g., domain-specific re-ranking) will further enhance its utility. This flexibility paves the way for broader application across multiple criteria decision-making environments, where relative comparisons between algorithm runs are essential even in the absence of a ground truth.

\subsection{Limitations and Future Directions} 
The \red{system} currently relies on pre-defined metrics and static visualization techniques, which may not capture all nuances of dynamic retrieval-generation interactions in rapidly evolving systems. Future work will focus on incorporating adaptive, plugin-based modules that allow for easy integration of emerging evolutionary operators and retrieval strategies. Furthermore, extensive user studies—spanning diverse expertise levels and application contexts—are necessary to validate and refine the \red{system's} design. By enhancing both the transparency of internal RAG operations and the interpretability of generated outputs, \sysname{} aims to provide a critical tool for both researchers and practitioners seeking to optimize the performance and trustworthiness of RAG-based systems.

% \subsection{Conclusion}
% In this paper, we present \sysname{}, a systematic evaluation \red{system} that supports users in diagnosing and optimizing RAG workflows. Drawing from a formative study (N=12) with RAG users, we uncovered key practices, challenges, and expectations in understanding retrieval relevance, tracing knowledge propagation, and resolving generation inconsistencies. Building on these insights, we developed a diagnostic methodology and interactive analysis tools that enable users to assess retrieval quality, identify generation errors, and refine retrieval strategies for improved RAG performance. A user study (N=11) and expert interviews demonstrated that \sysname{} could effectively facilitate troubleshooting, enhance users' understanding of retrieval-generation interactions, and foster more reliable and controllable RAG systems.

\section*{Acknowledgements}
This work was supported in part by the National Natural Science Foundation of China (No. 62202217), Guangdong Basic and Applied Basic Research Foundation (No. 2023A1515012889), and Guangdong Key Program (No. 2021QN02X794). An implementation of \sysname{} is available at \url{https://github.com/VIS-SUSTech/RAGTrace}.

\bibliographystyle{ACM-Reference-Format}
\bibliography{references}


\begin{thebibliography}{74}

%%% the Web address.

\ifx \showCODEN    \undefined \def \showCODEN     #1{\unskip}     \fi
\ifx \showISBNx    \undefined \def \showISBNx     #1{\unskip}     \fi
\ifx \showISBNxiii \undefined \def \showISBNxiii  #1{\unskip}     \fi
\ifx \showISSN     \undefined \def \showISSN      #1{\unskip}     \fi
\ifx \showLCCN     \undefined \def \showLCCN      #1{\unskip}     \fi
\ifx \shownote     \undefined \def \shownote      #1{#1}          \fi
\ifx \showarticletitle \undefined \def \showarticletitle #1{#1}   \fi
\ifx \showURL      \undefined \def \showURL       {\relax}        \fi
% invisible to TeX
\providecommand\bibfield[2]{#2}
\providecommand\bibinfo[2]{#2}
\providecommand\natexlab[1]{#1}
\providecommand\showeprint[2][]{arXiv:#2}

\bibitem[Abu-Rasheed et~al\mbox{.}(2024)]%
        {AbuRasheed2024Knowledge}
\bibfield{author}{\bibinfo{person}{Hasan Abu-Rasheed},
  \bibinfo{person}{Christian Weber}, {and} \bibinfo{person}{Madjid Fathi}.}
  \bibinfo{year}{2024}\natexlab{}.
\newblock \showarticletitle{Knowledge Graphs as Context Sources for LLM-Based
  Explanations of Learning Recommendations}. In
  \bibinfo{booktitle}{\emph{Proceedings of IEEE Global Engineering Education
  Conference}}. \bibinfo{publisher}{IEEE}, \bibinfo{address}{Kos Island,
  Greece}, \bibinfo{pages}{1--5}.
\newblock
\href{https://doi.org/10.1109/EDUCON60312.2024.10578654}{doi:\nolinkurl{10.1109/EDUCON60312.2024.10578654}}


\bibitem[Ahn and Brusilovsky(2013)]%
        {ahn2013adaptive}
\bibfield{author}{\bibinfo{person}{Jae-wook Ahn} {and} \bibinfo{person}{Peter
  Brusilovsky}.} \bibinfo{year}{2013}\natexlab{}.
\newblock \showarticletitle{Adaptive visualization for exploratory information
  retrieval}.
\newblock \bibinfo{journal}{\emph{Information Processing \& Management}}
  \bibinfo{volume}{49}, \bibinfo{number}{5} (\bibinfo{year}{2013}),
  \bibinfo{pages}{1139--1164}.
\newblock


\bibitem[Alicioglu and Sun(2022)]%
        {Gulsum2022survey}
\bibfield{author}{\bibinfo{person}{Gulsum Alicioglu} {and} \bibinfo{person}{Bo
  Sun}.} \bibinfo{year}{2022}\natexlab{}.
\newblock \showarticletitle{A survey of visual analytics for Explainable
  Artificial Intelligence methods}.
\newblock \bibinfo{journal}{\emph{Computers \& Graphics}}
  \bibinfo{volume}{102} (\bibinfo{year}{2022}), \bibinfo{pages}{502--520}.
\newblock
\showISSN{0097-8493}
\href{https://doi.org/10.1016/j.cag.2021.09.002}{doi:\nolinkurl{10.1016/j.cag.2021.09.002}}


\bibitem[Arawjo et~al\mbox{.}(2024)]%
        {Arawjo2024chi}
\bibfield{author}{\bibinfo{person}{Ian Arawjo}, \bibinfo{person}{Chelse
  Swoopes}, \bibinfo{person}{Priyan Vaithilingam}, \bibinfo{person}{Martin
  Wattenberg}, {and} \bibinfo{person}{Elena~L. Glassman}.}
  \bibinfo{year}{2024}\natexlab{}.
\newblock \showarticletitle{ChainForge: A Visual Toolkit for Prompt Engineering
  and LLM Hypothesis Testing}. In \bibinfo{booktitle}{\emph{Proceedings of the
  CHI Conference on Human Factors in Computing Systems}}.
  \bibinfo{publisher}{Association for Computing Machinery},
  \bibinfo{address}{New York, NY, USA}, \bibinfo{pages}{1--18}.
\newblock
\showISBNx{9798400703300}
\href{https://doi.org/10.1145/3613904.3642016}{doi:\nolinkurl{10.1145/3613904.3642016}}


\bibitem[Bahr et~al\mbox{.}(2025)]%
        {Lukas2025Knowledge}
\bibfield{author}{\bibinfo{person}{Lukas Bahr}, \bibinfo{person}{Christoph
  Wehner}, \bibinfo{person}{Judith Wewerka}, \bibinfo{person}{José
  Bittencourt}, \bibinfo{person}{Ute Schmid}, {and} \bibinfo{person}{Rüdiger
  Daub}.} \bibinfo{year}{2025}\natexlab{}.
\newblock \showarticletitle{Knowledge graph enhanced retrieval-augmented
  generation for failure mode and effects analysis}.
\newblock \bibinfo{journal}{\emph{Journal of Industrial Information
  Integration}}  \bibinfo{volume}{45} (\bibinfo{year}{2025}),
  \bibinfo{pages}{100807}.
\newblock
\showISSN{2452-414X}
\href{https://doi.org/10.1016/j.jii.2025.100807}{doi:\nolinkurl{10.1016/j.jii.2025.100807}}


\bibitem[Barnett et~al\mbox{.}(2024)]%
        {barnett2024seven}
\bibfield{author}{\bibinfo{person}{Scott Barnett}, \bibinfo{person}{Stefanus
  Kurniawan}, \bibinfo{person}{Srikanth Thudumu}, \bibinfo{person}{Zach
  Brannelly}, {and} \bibinfo{person}{Mohamed Abdelrazek}.}
  \bibinfo{year}{2024}\natexlab{}.
\newblock \showarticletitle{Seven Failure Points When Engineering a Retrieval
  Augmented Generation System}. In \bibinfo{booktitle}{\emph{Proceedings of the
  IEEE/ACM 3rd International Conference on AI Engineering - Software
  Engineering for AI}}. \bibinfo{publisher}{Association for Computing
  Machinery}, \bibinfo{address}{New York, NY, USA}, \bibinfo{pages}{194–199}.
\newblock
\showISBNx{9798400705915}
\href{https://doi.org/10.1145/3644815.3644945}{doi:\nolinkurl{10.1145/3644815.3644945}}


\bibitem[Belkin et~al\mbox{.}(1993)]%
        {belkin1993braque}
\bibfield{author}{\bibinfo{person}{Nicholas~J. Belkin},
  \bibinfo{person}{Pier~Giorgio Marchetti}, {and} \bibinfo{person}{Colleen
  Cool}.} \bibinfo{year}{1993}\natexlab{}.
\newblock \showarticletitle{BRAQUE: Design of an interface to support user
  interaction in information retrieval}.
\newblock \bibinfo{journal}{\emph{Information processing \& management}}
  \bibinfo{volume}{29}, \bibinfo{number}{3} (\bibinfo{year}{1993}),
  \bibinfo{pages}{325--344}.
\newblock


\bibitem[Bevilacqua et~al\mbox{.}(2025)]%
        {Bevilacqua2025Automated}
\bibfield{author}{\bibinfo{person}{Marialena Bevilacqua},
  \bibinfo{person}{Kezia Oketch}, \bibinfo{person}{Ruiyang Qin},
  \bibinfo{person}{Will Stamey}, \bibinfo{person}{Xinyuan Zhang},
  \bibinfo{person}{Yi Gan}, \bibinfo{person}{Kai Yang}, {and}
  \bibinfo{person}{Ahmed Abbasi}.} \bibinfo{year}{2025}\natexlab{}.
\newblock \showarticletitle{When Automated Assessment Meets Automated Content
  Generation: Examining Text Quality in the Era of GPTs}.
\newblock \bibinfo{journal}{\emph{ACM Transactions on Information Systems}}
  \bibinfo{volume}{43}, \bibinfo{number}{2} (\bibinfo{year}{2025}),
  \bibinfo{pages}{1--36}.
\newblock
\href{https://doi.org/10.1145/3702639}{doi:\nolinkurl{10.1145/3702639}}


\bibitem[Brasoveanu et~al\mbox{.}(2024)]%
        {Brasoveanu2024Visualizing}
\bibfield{author}{\bibinfo{person}{Adrian~M.P. Brasoveanu},
  \bibinfo{person}{Arno Scharl}, \bibinfo{person}{Lyndon~J.B. Nixon}, {and}
  \bibinfo{person}{Răzvan Andonie}.} \bibinfo{year}{2024}\natexlab{}.
\newblock \showarticletitle{Visualizing Large Language Models: A Brief Survey}.
  In \bibinfo{booktitle}{\emph{Proceedings of the International Conference
  Information Visualisation}}. \bibinfo{publisher}{IEEE},
  \bibinfo{address}{Coimbra, Portugal}, \bibinfo{pages}{236--245}.
\newblock
\href{https://doi.org/10.1109/IV64223.2024.00049}{doi:\nolinkurl{10.1109/IV64223.2024.00049}}


\bibitem[Chang et~al\mbox{.}(2024)]%
        {Chang2024ASurvey}
\bibfield{author}{\bibinfo{person}{Yupeng Chang}, \bibinfo{person}{Xu Wang},
  \bibinfo{person}{Jindong Wang}, \bibinfo{person}{Yuan Wu},
  \bibinfo{person}{Linyi Yang}, \bibinfo{person}{Kaijie Zhu},
  \bibinfo{person}{Hao Chen}, \bibinfo{person}{Xiaoyuan Yi},
  \bibinfo{person}{Cunxiang Wang}, \bibinfo{person}{Yidong Wang},
  \bibinfo{person}{Wei Ye}, \bibinfo{person}{Yue Zhang}, \bibinfo{person}{Yi
  Chang}, \bibinfo{person}{Philip~S. Yu}, \bibinfo{person}{Qiang Yang}, {and}
  \bibinfo{person}{Xing Xie}.} \bibinfo{year}{2024}\natexlab{}.
\newblock \showarticletitle{A Survey on Evaluation of Large Language Models}.
\newblock \bibinfo{journal}{\emph{ACM Transactions on Intelligent Systems and
  Technology}} \bibinfo{volume}{15}, \bibinfo{number}{3}
  (\bibinfo{year}{2024}), \bibinfo{numpages}{45}~pages.
\newblock
\showISSN{2157-6904}
\href{https://doi.org/10.1145/3641289}{doi:\nolinkurl{10.1145/3641289}}


\bibitem[Chaudhuri and Narasayya(1998)]%
        {chaudhuri1998autoadmin}
\bibfield{author}{\bibinfo{person}{Surajit Chaudhuri} {and}
  \bibinfo{person}{Vivek Narasayya}.} \bibinfo{year}{1998}\natexlab{}.
\newblock \showarticletitle{AutoAdmin “what-if” index analysis utility}.
\newblock \bibinfo{journal}{\emph{ACM SIGMOD Record}} \bibinfo{volume}{27},
  \bibinfo{number}{2} (\bibinfo{year}{1998}), \bibinfo{pages}{367--378}.
\newblock


\bibitem[Chen et~al\mbox{.}(2024)]%
        {chen2024benchmarking}
\bibfield{author}{\bibinfo{person}{Jiawei Chen}, \bibinfo{person}{Hongyu Lin},
  \bibinfo{person}{Xianpei Han}, {and} \bibinfo{person}{Le Sun}.}
  \bibinfo{year}{2024}\natexlab{}.
\newblock \showarticletitle{Benchmarking large language models in
  retrieval-augmented generation}. In \bibinfo{booktitle}{\emph{Proceedings of
  the AAAI Conference on Artificial Intelligence}}. \bibinfo{publisher}{AAAI
  Press}, \bibinfo{address}{Vancouver, Canada}, \bibinfo{pages}{17754--17762}.
\newblock
\href{https://doi.org/10.1609/aaai.v38i16.29728}{doi:\nolinkurl{10.1609/aaai.v38i16.29728}}


\bibitem[Cheng et~al\mbox{.}(2024)]%
        {cheng2024relic}
\bibfield{author}{\bibinfo{person}{Furui Cheng}, \bibinfo{person}{Vil\'{e}m
  Zouhar}, \bibinfo{person}{Simran Arora}, \bibinfo{person}{Mrinmaya Sachan},
  \bibinfo{person}{Hendrik Strobelt}, {and} \bibinfo{person}{Mennatallah
  El-Assady}.} \bibinfo{year}{2024}\natexlab{}.
\newblock \showarticletitle{RELIC: Investigating Large Language Model Responses
  using Self-Consistency}. In \bibinfo{booktitle}{\emph{Proceedings of the CHI
  Conference on Human Factors in Computing Systems}}.
  \bibinfo{publisher}{Association for Computing Machinery},
  \bibinfo{address}{New York, NY, USA}, \bibinfo{pages}{1--18}.
\newblock
\showISBNx{9798400703300}
\href{https://doi.org/10.1145/3613904.3641904}{doi:\nolinkurl{10.1145/3613904.3641904}}


\bibitem[Coscia and Endert(2024)]%
        {Coscia2024KnowledgeVIS}
\bibfield{author}{\bibinfo{person}{Adam Coscia} {and} \bibinfo{person}{Alex
  Endert}.} \bibinfo{year}{2024}\natexlab{}.
\newblock \showarticletitle{KnowledgeVIS: Interpreting Language Models by
  Comparing Fill-in-the-Blank Prompts}.
\newblock \bibinfo{journal}{\emph{IEEE Transactions on Visualization and
  Computer Graphics}} \bibinfo{volume}{30}, \bibinfo{number}{9}
  (\bibinfo{year}{2024}), \bibinfo{pages}{6520–6532}.
\newblock
\showISSN{1077-2626}
\href{https://doi.org/10.1109/TVCG.2023.3346713}{doi:\nolinkurl{10.1109/TVCG.2023.3346713}}


\bibitem[Coscia et~al\mbox{.}(2024)]%
        {coscia2024iscore}
\bibfield{author}{\bibinfo{person}{Adam Coscia}, \bibinfo{person}{Langdon
  Holmes}, \bibinfo{person}{Wesley Morris}, \bibinfo{person}{Joon~Suh Choi},
  \bibinfo{person}{Scott Crossley}, {and} \bibinfo{person}{Alex Endert}.}
  \bibinfo{year}{2024}\natexlab{}.
\newblock \showarticletitle{i{S}core: Visual Analytics for Interpreting How
  Language Models Automatically Score Summaries}. In
  \bibinfo{booktitle}{\emph{Proceedings of the International Conference on
  Intelligent User Interfaces}}. \bibinfo{publisher}{Association for Computing
  Machinery}, \bibinfo{address}{New York, NY, USA}, \bibinfo{pages}{787--802}.
\newblock
\showISBNx{9798400705083}
\href{https://doi.org/10.1145/3640543.3645142}{doi:\nolinkurl{10.1145/3640543.3645142}}


\bibitem[Cuconasu et~al\mbox{.}(2024)]%
        {Cuconasu2024Power}
\bibfield{author}{\bibinfo{person}{Florin Cuconasu}, \bibinfo{person}{Giovanni
  Trappolini}, \bibinfo{person}{Federico Siciliano}, \bibinfo{person}{Simone
  Filice}, \bibinfo{person}{Cesare Campagnano}, \bibinfo{person}{Yoelle
  Maarek}, \bibinfo{person}{Nicola Tonellotto}, {and} \bibinfo{person}{Fabrizio
  Silvestri}.} \bibinfo{year}{2024}\natexlab{}.
\newblock \showarticletitle{The Power of Noise: Redefining Retrieval for RAG
  Systems}. In \bibinfo{booktitle}{\emph{Proceedings of the International ACM
  SIGIR Conference on Research and Development in Information Retrieval}}.
  \bibinfo{publisher}{Association for Computing Machinery},
  \bibinfo{address}{New York, NY, USA}, \bibinfo{pages}{719--729}.
\newblock
\showISBNx{9798400704314}
\href{https://doi.org/10.1145/3626772.3657834}{doi:\nolinkurl{10.1145/3626772.3657834}}


\bibitem[Datta et~al\mbox{.}(2022)]%
        {pmlr-v176-datta22a}
\bibfield{author}{\bibinfo{person}{Anupam Datta}, \bibinfo{person}{Matt
  Fredrikson}, \bibinfo{person}{Klas Leino}, \bibinfo{person}{Kaiji Lu},
  \bibinfo{person}{Shayak Sen}, \bibinfo{person}{Ricardo Shih}, {and}
  \bibinfo{person}{Zifan Wang}.} \bibinfo{year}{2022}\natexlab{}.
\newblock \showarticletitle{Exploring Conceptual Soundness with TruLens}. In
  \bibinfo{booktitle}{\emph{Proceedings of the NeurIPS Competitions and
  Demonstrations Track}}, Vol.~\bibinfo{volume}{176}.
  \bibinfo{publisher}{PMLR}, \bibinfo{pages}{302--307}.
\newblock


\bibitem[{Edmond Ghali} et~al\mbox{.}(2024)]%
        {Julien2024Enhancing}
\bibfield{author}{\bibinfo{person}{Julien~Pierre {Edmond Ghali}},
  \bibinfo{person}{Kosuke Shima}, \bibinfo{person}{Koichi Moriyama},
  \bibinfo{person}{Atsuko Mutoh}, {and} \bibinfo{person}{Nobuhiro Inuzuka}.}
  \bibinfo{year}{2024}\natexlab{}.
\newblock \showarticletitle{Enhancing Retrieval Processes for Language
  Generation with Augmented Queries to Provide Factual Information on
  Schizophrenia}.
\newblock \bibinfo{journal}{\emph{Procedia Computer Science}}
  \bibinfo{volume}{246} (\bibinfo{year}{2024}), \bibinfo{pages}{443--452}.
\newblock
\showISSN{1877-0509}
\href{https://doi.org/10.1016/j.procs.2024.09.424}{doi:\nolinkurl{10.1016/j.procs.2024.09.424}}


\bibitem[Ellis(1989)]%
        {ellis1989behavioural}
\bibfield{author}{\bibinfo{person}{David Ellis}.}
  \bibinfo{year}{1989}\natexlab{}.
\newblock \showarticletitle{A behavioural approach to information retrieval
  system design}.
\newblock \bibinfo{journal}{\emph{Journal of documentation}}
  \bibinfo{volume}{45}, \bibinfo{number}{3} (\bibinfo{year}{1989}),
  \bibinfo{pages}{171--212}.
\newblock


\bibitem[Es et~al\mbox{.}(2024)]%
        {es-etal-2024-ragas}
\bibfield{author}{\bibinfo{person}{Shahul Es}, \bibinfo{person}{Jithin James},
  \bibinfo{person}{Luis Espinosa~Anke}, {and} \bibinfo{person}{Steven
  Schockaert}.} \bibinfo{year}{2024}\natexlab{}.
\newblock \showarticletitle{{RAGA}s: Automated Evaluation of Retrieval
  Augmented Generation}. In \bibinfo{booktitle}{\emph{Proceedings of the 18th
  Conference of the European Chapter of the Association for Computational
  Linguistics: System Demonstrations}}. \bibinfo{publisher}{Association for
  Computational Linguistics}, \bibinfo{address}{St. Julians, Malta},
  \bibinfo{pages}{150--158}.
\newblock


\bibitem[Fan et~al\mbox{.}(2024a)]%
        {fan2024survey}
\bibfield{author}{\bibinfo{person}{Wenqi Fan}, \bibinfo{person}{Yujuan Ding},
  \bibinfo{person}{Liangbo Ning}, \bibinfo{person}{Shijie Wang},
  \bibinfo{person}{Hengyun Li}, \bibinfo{person}{Dawei Yin},
  \bibinfo{person}{Tat-Seng Chua}, {and} \bibinfo{person}{Qing Li}.}
  \bibinfo{year}{2024}\natexlab{a}.
\newblock \showarticletitle{A Survey on RAG Meeting LLMs: Towards
  Retrieval-Augmented Large Language Models}. In
  \bibinfo{booktitle}{\emph{Proceedings of the ACM SIGKDD Conference on
  Knowledge Discovery and Data Mining}}. \bibinfo{publisher}{Association for
  Computing Machinery}, \bibinfo{address}{New York, NY, USA},
  \bibinfo{pages}{6491--6501}.
\newblock
\showISBNx{9798400704901}
\href{https://doi.org/10.1145/3637528.3671470}{doi:\nolinkurl{10.1145/3637528.3671470}}


\bibitem[Fan et~al\mbox{.}(2024b)]%
        {ragsurvey}
\bibfield{author}{\bibinfo{person}{Wenqi Fan}, \bibinfo{person}{Yujuan Ding},
  \bibinfo{person}{Liangbo Ning}, \bibinfo{person}{Shijie Wang},
  \bibinfo{person}{Hengyun Li}, \bibinfo{person}{Dawei Yin},
  \bibinfo{person}{Tat-Seng Chua}, {and} \bibinfo{person}{Qing Li}.}
  \bibinfo{year}{2024}\natexlab{b}.
\newblock \showarticletitle{A Survey on RAG Meeting LLMs: Towards
  Retrieval-Augmented Large Language Models}. In
  \bibinfo{booktitle}{\emph{Proceedings of the 30th ACM SIGKDD Conference on
  Knowledge Discovery and Data Mining}}. \bibinfo{publisher}{Association for
  Computing Machinery}, \bibinfo{address}{New York, NY, USA},
  \bibinfo{pages}{6491–6501}.
\newblock
\showISBNx{9798400704901}
\href{https://doi.org/10.1145/3637528.3671470}{doi:\nolinkurl{10.1145/3637528.3671470}}


\bibitem[Fu et~al\mbox{.}(2024)]%
        {Fu2024uist}
\bibfield{author}{\bibinfo{person}{Yu Fu}, \bibinfo{person}{Shunan Guo},
  \bibinfo{person}{Jane Hoffswell}, \bibinfo{person}{Victor S.~Bursztyn},
  \bibinfo{person}{Ryan Rossi}, {and} \bibinfo{person}{John Stasko}.}
  \bibinfo{year}{2024}\natexlab{}.
\newblock \showarticletitle{{"The Data Says Otherwise" — Towards Automated
  Fact-Checking and Communication of Data Claims}}. In
  \bibinfo{booktitle}{\emph{Proceedings of the Annual ACM Symposium on User
  Interface Software and Technology}}. \bibinfo{publisher}{Association for
  Computing Machinery}, \bibinfo{address}{New York, NY, USA},
  \bibinfo{pages}{1--20}.
\newblock
\showISBNx{9798400706288}
\href{https://doi.org/10.1145/3654777.3676359}{doi:\nolinkurl{10.1145/3654777.3676359}}


\bibitem[Gao et~al\mbox{.}(2023)]%
        {gao-etal-2023-precise}
\bibfield{author}{\bibinfo{person}{Luyu Gao}, \bibinfo{person}{Xueguang Ma},
  \bibinfo{person}{Jimmy Lin}, {and} \bibinfo{person}{Jamie Callan}.}
  \bibinfo{year}{2023}\natexlab{}.
\newblock \showarticletitle{Precise Zero-Shot Dense Retrieval without Relevance
  Labels}. In \bibinfo{booktitle}{\emph{Proceedings of the 61st Annual Meeting
  of the Association for Computational Linguistics (Volume 1: Long Papers)}}.
  \bibinfo{publisher}{Association for Computational Linguistics},
  \bibinfo{address}{Toronto, Canada}, \bibinfo{pages}{1762--1777}.
\newblock
\href{https://doi.org/v1/2023.acl-long.99}{doi:\nolinkurl{v1/2023.acl-long.99}}


\bibitem[Han et~al\mbox{.}(2024)]%
        {han2024automating}
\bibfield{author}{\bibinfo{person}{Binglan Han}, \bibinfo{person}{Teo Susnjak},
  {and} \bibinfo{person}{Anuradha Mathrani}.} \bibinfo{year}{2024}\natexlab{}.
\newblock \showarticletitle{Automating Systematic Literature Reviews with
  Retrieval-Augmented Generation: A Comprehensive Overview}.
\newblock \bibinfo{journal}{\emph{Applied Sciences}} \bibinfo{volume}{14},
  \bibinfo{number}{19} (\bibinfo{year}{2024}), \bibinfo{numpages}{17}~pages.
\newblock
\showISSN{2076-3417}
\href{https://doi.org/10.3390/app14199103}{doi:\nolinkurl{10.3390/app14199103}}


\bibitem[Hassija et~al\mbox{.}(2024)]%
        {hassija2024interpreting}
\bibfield{author}{\bibinfo{person}{Vikas Hassija}, \bibinfo{person}{Vinay
  Chamola}, \bibinfo{person}{Atmesh Mahapatra}, \bibinfo{person}{Abhinandan
  Singal}, \bibinfo{person}{Divyansh Goel}, \bibinfo{person}{Kaizhu Huang},
  \bibinfo{person}{Simone Scardapane}, \bibinfo{person}{Indro Spinelli},
  \bibinfo{person}{Mufti Mahmud}, {and} \bibinfo{person}{Amir Hussain}.}
  \bibinfo{year}{2024}\natexlab{}.
\newblock \showarticletitle{Interpreting Black-Box Models: A Review on
  Explainable Artificial Intelligence}.
\newblock \bibinfo{journal}{\emph{Cognitive Computation}} \bibinfo{volume}{16},
  \bibinfo{number}{1} (\bibinfo{year}{2024}), \bibinfo{pages}{45--74}.
\newblock
\href{https://doi.org/10.1007/s12559-023-10179-8}{doi:\nolinkurl{10.1007/s12559-023-10179-8}}


\bibitem[Hoque et~al\mbox{.}(2024)]%
        {Hoque2024HaLLMark}
\bibfield{author}{\bibinfo{person}{Md~Naimul Hoque}, \bibinfo{person}{Tasfia
  Mashiat}, \bibinfo{person}{Bhavya Ghai}, \bibinfo{person}{Cecilia~D.
  Shelton}, \bibinfo{person}{Fanny Chevalier}, \bibinfo{person}{Kari Kraus},
  {and} \bibinfo{person}{Niklas Elmqvist}.} \bibinfo{year}{2024}\natexlab{}.
\newblock \showarticletitle{The HaLLMark Effect: Supporting Provenance and
  Transparent Use of Large Language Models in Writing with Interactive
  Visualization}. In \bibinfo{booktitle}{\emph{Proceedings of the CHI
  Conference on Human Factors in Computing Systems}}.
  \bibinfo{publisher}{Association for Computing Machinery},
  \bibinfo{address}{New York, NY, USA}, \bibinfo{pages}{1--15}.
\newblock
\showISBNx{9798400703300}
\href{https://doi.org/10.1145/3613904.3641895}{doi:\nolinkurl{10.1145/3613904.3641895}}


\bibitem[Ingwersen(1992)]%
        {ingwersen1992information}
\bibfield{author}{\bibinfo{person}{Peter Ingwersen}.}
  \bibinfo{year}{1992}\natexlab{}.
\newblock \bibinfo{booktitle}{\emph{Information retrieval interaction}}.
  Vol.~\bibinfo{volume}{246}.
\newblock \bibinfo{publisher}{Taylor Graham Publishing},
  \bibinfo{address}{GBR}.
\newblock
\showISBNx{0947568549}


\bibitem[Kahng et~al\mbox{.}(2024)]%
        {Kahng2024chi}
\bibfield{author}{\bibinfo{person}{Minsuk Kahng}, \bibinfo{person}{Ian Tenney},
  \bibinfo{person}{Mahima Pushkarna}, \bibinfo{person}{Michael~Xieyang Liu},
  \bibinfo{person}{James Wexler}, \bibinfo{person}{Emily Reif},
  \bibinfo{person}{Krystal Kallarackal}, \bibinfo{person}{Minsuk Chang},
  \bibinfo{person}{Michael Terry}, {and} \bibinfo{person}{Lucas Dixon}.}
  \bibinfo{year}{2024}\natexlab{}.
\newblock \showarticletitle{LLM Comparator: Visual Analytics for Side-by-Side
  Evaluation of Large Language Models}. In \bibinfo{booktitle}{\emph{Extended
  Abstracts of the CHI Conference on Human Factors in Computing Systems}}.
  \bibinfo{publisher}{Association for Computing Machinery},
  \bibinfo{address}{New York, NY, USA}, \bibinfo{pages}{1--7}.
\newblock
\showISBNx{9798400703317}
\href{https://doi.org/10.1145/3613905.3650755}{doi:\nolinkurl{10.1145/3613905.3650755}}


\bibitem[Kahng et~al\mbox{.}(2025)]%
        {Kahng2025Comparator}
\bibfield{author}{\bibinfo{person}{Minsuk Kahng}, \bibinfo{person}{Ian Tenney},
  \bibinfo{person}{Mahima Pushkarna}, \bibinfo{person}{Michael~Xieyang Liu},
  \bibinfo{person}{James Wexler}, \bibinfo{person}{Emily Reif},
  \bibinfo{person}{Krystal Kallarackal}, \bibinfo{person}{Minsuk Chang},
  \bibinfo{person}{Michael Terry}, {and} \bibinfo{person}{Lucas Dixon}.}
  \bibinfo{year}{2025}\natexlab{}.
\newblock \showarticletitle{LLM Comparator: Interactive Analysis of
  Side-by-Side Evaluation of Large Language Models}.
\newblock \bibinfo{journal}{\emph{IEEE Transactions on Visualization and
  Computer Graphics}} \bibinfo{volume}{31}, \bibinfo{number}{1}
  (\bibinfo{year}{2025}), \bibinfo{pages}{503--513}.
\newblock
\href{https://doi.org/10.1109/TVCG.2024.3456354}{doi:\nolinkurl{10.1109/TVCG.2024.3456354}}


\bibitem[Kamalloo et~al\mbox{.}(2023)]%
        {kamalloo-etal-2023-evaluating}
\bibfield{author}{\bibinfo{person}{Ehsan Kamalloo}, \bibinfo{person}{Nouha
  Dziri}, \bibinfo{person}{Charles Clarke}, {and} \bibinfo{person}{Davood
  Rafiei}.} \bibinfo{year}{2023}\natexlab{}.
\newblock \showarticletitle{Evaluating Open-Domain Question Answering in the
  Era of Large Language Models}. In \bibinfo{booktitle}{\emph{Proceedings of
  the Annual Meeting of the Association for Computational Linguistics}}.
  \bibinfo{publisher}{Association for Computational Linguistics},
  \bibinfo{address}{Toronto, Canada}, \bibinfo{pages}{5591--5606}.
\newblock
\href{https://doi.org/10.18653/v1/2023.acl-long.307}{doi:\nolinkurl{10.18653/v1/2023.acl-long.307}}


\bibitem[Kim et~al\mbox{.}(2024)]%
        {Kim2024chi}
\bibfield{author}{\bibinfo{person}{Tae~Soo Kim}, \bibinfo{person}{Yoonjoo Lee},
  \bibinfo{person}{Jamin Shin}, \bibinfo{person}{Young-Ho Kim}, {and}
  \bibinfo{person}{Juho Kim}.} \bibinfo{year}{2024}\natexlab{}.
\newblock \showarticletitle{EvalLM: Interactive Evaluation of Large Language
  Model Prompts on User-Defined Criteria}. In
  \bibinfo{booktitle}{\emph{Proceedings of the CHI Conference on Human Factors
  in Computing Systems}}. \bibinfo{publisher}{Association for Computing
  Machinery}, \bibinfo{address}{New York, NY, USA}, \bibinfo{pages}{1--21}.
\newblock
\showISBNx{9798400703300}
\href{https://doi.org/10.1145/3613904.3642216}{doi:\nolinkurl{10.1145/3613904.3642216}}


\bibitem[Kwiatkowski et~al\mbox{.}(2019)]%
        {nq}
\bibfield{author}{\bibinfo{person}{Tom Kwiatkowski},
  \bibinfo{person}{Jennimaria Palomaki}, \bibinfo{person}{Olivia Redfield},
  \bibinfo{person}{Michael Collins}, \bibinfo{person}{Ankur Parikh},
  \bibinfo{person}{Chris Alberti}, \bibinfo{person}{Danielle Epstein},
  \bibinfo{person}{Illia Polosukhin}, \bibinfo{person}{Matthew Kelcey},
  \bibinfo{person}{Jacob Devlin}, \bibinfo{person}{Kenton Lee},
  \bibinfo{person}{Kristina~N. Toutanova}, \bibinfo{person}{Llion Jones},
  \bibinfo{person}{Ming-Wei Chang}, \bibinfo{person}{Andrew Dai},
  \bibinfo{person}{Jakob Uszkoreit}, \bibinfo{person}{Quoc Le}, {and}
  \bibinfo{person}{Slav Petrov}.} \bibinfo{year}{2019}\natexlab{}.
\newblock \showarticletitle{Natural Questions: a Benchmark for Question
  Answering Research}.
\newblock \bibinfo{journal}{\emph{Transactions of the Association of
  Computational Linguistics}}  \bibinfo{volume}{7} (\bibinfo{year}{2019}),
  \bibinfo{pages}{452--466}.
\newblock
\href{https://doi.org/10.1162/tacl_a_00276}{doi:\nolinkurl{10.1162/tacl_a_00276}}


\bibitem[La~Rosa et~al\mbox{.}(2023)]%
        {La2023State}
\bibfield{author}{\bibinfo{person}{B. La~Rosa}, \bibinfo{person}{G. Blasilli},
  \bibinfo{person}{R. Bourqui}, \bibinfo{person}{D. Auber}, \bibinfo{person}{G.
  Santucci}, \bibinfo{person}{R. Capobianco}, \bibinfo{person}{E. Bertini},
  \bibinfo{person}{R. Giot}, {and} \bibinfo{person}{M. Angelini}.}
  \bibinfo{year}{2023}\natexlab{}.
\newblock \showarticletitle{State of the Art of Visual Analytics for
  eXplainable Deep Learning}.
\newblock \bibinfo{journal}{\emph{Computer Graphics Forum}}
  \bibinfo{volume}{42}, \bibinfo{number}{1} (\bibinfo{year}{2023}),
  \bibinfo{pages}{319--355}.
\newblock
\showeprint{https://doi.org/10.1111/cgf.14733}


\bibitem[Lee et~al\mbox{.}(2025)]%
        {yu2024towards}
\bibfield{author}{\bibinfo{person}{Sam Yu-Te Lee}, \bibinfo{person}{Aryaman
  Bahukhandi}, \bibinfo{person}{Dongyu Liu}, {and} \bibinfo{person}{Kwan-Liu
  Ma}.} \bibinfo{year}{2025}\natexlab{}.
\newblock \showarticletitle{Towards Dataset-Scale and Feature-Oriented
  Evaluation of Text Summarization in Large Language Model Prompts}.
\newblock \bibinfo{journal}{\emph{IEEE Transactions on Visualization and
  Computer Graphics}} \bibinfo{volume}{31}, \bibinfo{number}{1}
  (\bibinfo{year}{2025}), \bibinfo{pages}{481--491}.
\newblock
\href{https://doi.org/10.1109/TVCG.2024.3456398}{doi:\nolinkurl{10.1109/TVCG.2024.3456398}}


\bibitem[Li et~al\mbox{.}(2024)]%
        {li2024linkq}
\bibfield{author}{\bibinfo{person}{Harry Li}, \bibinfo{person}{Gabriel
  Appleby}, {and} \bibinfo{person}{Ashley Suh}.}
  \bibinfo{year}{2024}\natexlab{}.
\newblock \showarticletitle{LinkQ: An LLM-Assisted Visual Interface for
  Knowledge Graph Question-Answering}. In \bibinfo{booktitle}{\emph{Proceedings
  of IEEE Visualization and Visual Analytics}}. \bibinfo{publisher}{IEEE},
  \bibinfo{address}{St. Pete Beach, FL, USA}, \bibinfo{pages}{116--120}.
\newblock
\href{https://doi.org/10.1109/VIS55277.2024.00031}{doi:\nolinkurl{10.1109/VIS55277.2024.00031}}


\bibitem[Li et~al\mbox{.}(2025)]%
        {Li2025Matching}
\bibfield{author}{\bibinfo{person}{Xiaoxi Li}, \bibinfo{person}{Jiajie Jin},
  \bibinfo{person}{Yujia Zhou}, \bibinfo{person}{Yuyao Zhang},
  \bibinfo{person}{Peitian Zhang}, \bibinfo{person}{Yutao Zhu}, {and}
  \bibinfo{person}{Zhicheng Dou}.} \bibinfo{year}{2025}\natexlab{}.
\newblock \showarticletitle{From Matching to Generation: A Survey on Generative
  Information Retrieval}.
\newblock \bibinfo{journal}{\emph{ACM Transactions on Information Systems}}
  (\bibinfo{year}{2025}).
\newblock
\showISSN{1046-8188}
\href{https://doi.org/10.1145/3722552}{doi:\nolinkurl{10.1145/3722552}}


\bibitem[Lyu et~al\mbox{.}(2025)]%
        {Lyu2025CRUDRAG}
\bibfield{author}{\bibinfo{person}{Yuanjie Lyu}, \bibinfo{person}{Zhiyu Li},
  \bibinfo{person}{Simin Niu}, \bibinfo{person}{Feiyu Xiong},
  \bibinfo{person}{Bo Tang}, \bibinfo{person}{Wenjin Wang},
  \bibinfo{person}{Hao Wu}, \bibinfo{person}{Huanyong Liu},
  \bibinfo{person}{Tong Xu}, {and} \bibinfo{person}{Enhong Chen}.}
  \bibinfo{year}{2025}\natexlab{}.
\newblock \showarticletitle{CRUD-RAG: A Comprehensive Chinese Benchmark for
  Retrieval-Augmented Generation of Large Language Models}.
\newblock \bibinfo{journal}{\emph{ACM Transactions on Information Systems}}
  \bibinfo{volume}{43}, \bibinfo{number}{2} (\bibinfo{year}{2025}),
  \bibinfo{pages}{359--369}.
\newblock
\showISSN{1046-8188}
\href{https://doi.org/10.1145/3701228}{doi:\nolinkurl{10.1145/3701228}}


\bibitem[Mansurova et~al\mbox{.}(2024)]%
        {mansurova2024qa}
\bibfield{author}{\bibinfo{person}{Aigerim Mansurova}, \bibinfo{person}{Aiganym
  Mansurova}, {and} \bibinfo{person}{Aliya Nugumanova}.}
  \bibinfo{year}{2024}\natexlab{}.
\newblock \showarticletitle{QA-RAG: Exploring LLM Reliance on External
  Knowledge}.
\newblock \bibinfo{journal}{\emph{Big Data and Cognitive Computing}}
  \bibinfo{volume}{8}, \bibinfo{number}{9} (\bibinfo{year}{2024}),
  \bibinfo{numpages}{15}~pages.
\newblock
\showISSN{2504-2289}
\href{https://doi.org/10.3390/bdcc8090115}{doi:\nolinkurl{10.3390/bdcc8090115}}


\bibitem[O'Hara and Fleger(2020)]%
        {osti_1644018}
\bibfield{author}{\bibinfo{person}{John~M. O'Hara} {and} \bibinfo{person}{S.
  Fleger}.} \bibinfo{year}{2020}\natexlab{}.
\newblock \bibinfo{booktitle}{\emph{Human-System Interface Design Review
  Guidelines}}.
\newblock \bibinfo{type}{{T}echnical {R}eport}.
  \bibinfo{institution}{Brookhaven National Lab. (BNL), Upton, NY (United
  States)}.
\newblock
\href{https://doi.org/1644018}{doi:\nolinkurl{1644018}}


\bibitem[Oral et~al\mbox{.}(2024)]%
        {oral2023information}
\bibfield{author}{\bibinfo{person}{Emre Oral}, \bibinfo{person}{Ria Chawla},
  \bibinfo{person}{Michel Wijkstra}, \bibinfo{person}{Narges Mahyar}, {and}
  \bibinfo{person}{Evanthia Dimara}.} \bibinfo{year}{2024}\natexlab{}.
\newblock \showarticletitle{From Information to Choice: A Critical Inquiry Into
  Visualization Tools for Decision Making}.
\newblock \bibinfo{journal}{\emph{IEEE Transactions on Visualization and
  Computer Graphics}} \bibinfo{volume}{30}, \bibinfo{number}{1}
  (\bibinfo{year}{2024}), \bibinfo{pages}{359--369}.
\newblock
\href{https://doi.org/10.1109/TVCG.2023.3326593}{doi:\nolinkurl{10.1109/TVCG.2023.3326593}}


\bibitem[Packowski et~al\mbox{.}(2025)]%
        {Packowski2025Optimizing}
\bibfield{author}{\bibinfo{person}{Sarah Packowski}, \bibinfo{person}{Inge
  Halilovic}, \bibinfo{person}{Jenifer Schlotfeldt}, {and}
  \bibinfo{person}{Trish Smith}.} \bibinfo{year}{2025}\natexlab{}.
\newblock \showarticletitle{Optimizing and Evaluating Enterprise
  Retrieval-Augmented Generation (RAG): A Content Design Perspective}. In
  \bibinfo{booktitle}{\emph{Proceedings of the International Conference on
  Advances in Artificial Intelligence}}. \bibinfo{publisher}{Association for
  Computing Machinery}, \bibinfo{address}{New York, NY, USA},
  \bibinfo{pages}{162--167}.
\newblock
\showISBNx{9798400718014}
\href{https://doi.org/10.1145/3704137.3704181}{doi:\nolinkurl{10.1145/3704137.3704181}}


\bibitem[Papineni et~al\mbox{.}(2002)]%
        {Papineni2002Proceedings}
\bibfield{author}{\bibinfo{person}{Kishore Papineni}, \bibinfo{person}{Salim
  Roukos}, \bibinfo{person}{Todd Ward}, {and} \bibinfo{person}{Wei-Jing Zhu}.}
  \bibinfo{year}{2002}\natexlab{}.
\newblock \showarticletitle{BLEU: a method for automatic evaluation of machine
  translation}. In \bibinfo{booktitle}{\emph{Proceedings of the Annual Meeting
  on Association for Computational Linguistics}}.
  \bibinfo{publisher}{Association for Computational Linguistics},
  \bibinfo{address}{USA}, \bibinfo{pages}{311--318}.
\newblock
\href{https://doi.org/10.3115/1073083.1073135}{doi:\nolinkurl{10.3115/1073083.1073135}}


\bibitem[Poli{\v c}ar et~al\mbox{.}(2024)]%
        {Policar2024}
\bibfield{author}{\bibinfo{person}{Pavlin~G. Poli{\v c}ar},
  \bibinfo{person}{Martin Stra{\v z}ar}, {and} \bibinfo{person}{Bla{\v z}
  Zupan}.} \bibinfo{year}{2024}\natexlab{}.
\newblock \showarticletitle{openTSNE: A Modular Python Library for t-SNE
  Dimensionality Reduction and Embedding}.
\newblock \bibinfo{journal}{\emph{Journal of Statistical Software}}
  \bibinfo{volume}{109}, \bibinfo{number}{3} (\bibinfo{year}{2024}),
  \bibinfo{pages}{1--30}.
\newblock
\href{https://doi.org/10.18637/jss.v109.i03}{doi:\nolinkurl{10.18637/jss.v109.i03}}


\bibitem[Raiaan et~al\mbox{.}(2024)]%
        {Raiaan2024Review}
\bibfield{author}{\bibinfo{person}{Mohaimenul Azam~Khan Raiaan},
  \bibinfo{person}{Md. Saddam~Hossain Mukta}, \bibinfo{person}{Kaniz Fatema},
  \bibinfo{person}{Nur~Mohammad Fahad}, \bibinfo{person}{Sadman Sakib},
  \bibinfo{person}{Most Marufatul~Jannat Mim}, \bibinfo{person}{Jubaer Ahmad},
  \bibinfo{person}{Mohammed~Eunus Ali}, {and} \bibinfo{person}{Sami Azam}.}
  \bibinfo{year}{2024}\natexlab{}.
\newblock \showarticletitle{A Review on Large Language Models: Architectures,
  Applications, Taxonomies, Open Issues and Challenges}.
\newblock \bibinfo{journal}{\emph{IEEE Access}}  \bibinfo{volume}{12}
  (\bibinfo{year}{2024}), \bibinfo{pages}{26839--26874}.
\newblock
\href{https://doi.org/10.1109/ACCESS.2024.3365742}{doi:\nolinkurl{10.1109/ACCESS.2024.3365742}}


\bibitem[Robertson et~al\mbox{.}(2023)]%
        {Robertson2023chi}
\bibfield{author}{\bibinfo{person}{Samantha Robertson},
  \bibinfo{person}{Zijie~J. Wang}, \bibinfo{person}{Dominik Moritz},
  \bibinfo{person}{Mary~Beth Kery}, {and} \bibinfo{person}{Fred Hohman}.}
  \bibinfo{year}{2023}\natexlab{}.
\newblock \showarticletitle{Angler: Helping Machine Translation Practitioners
  Prioritize Model Improvements}. In \bibinfo{booktitle}{\emph{Proceedings of
  the CHI Conference on Human Factors in Computing Systems}}.
  \bibinfo{publisher}{Association for Computing Machinery},
  \bibinfo{address}{New York, NY, USA}, \bibinfo{pages}{1--20}.
\newblock
\showISBNx{9781450394215}
\href{https://doi.org/10.1145/3544548.3580790}{doi:\nolinkurl{10.1145/3544548.3580790}}


\bibitem[Ru et~al\mbox{.}(2024)]%
        {ru2024ragchecker}
\bibfield{author}{\bibinfo{person}{Dongyu Ru}, \bibinfo{person}{Lin Qiu},
  \bibinfo{person}{Xiangkun Hu}, \bibinfo{person}{Tianhang Zhang},
  \bibinfo{person}{Peng Shi}, \bibinfo{person}{Shuaichen Chang},
  \bibinfo{person}{Cheng Jiayang}, \bibinfo{person}{Cunxiang Wang},
  \bibinfo{person}{Shichao Sun}, \bibinfo{person}{Huanyu Li}, {et~al\mbox{.}}}
  \bibinfo{year}{2024}\natexlab{}.
\newblock \showarticletitle{Ragchecker: A fine-grained framework for diagnosing
  retrieval-augmented generation}.
\newblock \bibinfo{journal}{\emph{Advances in Neural Information Processing
  Systems}}  \bibinfo{volume}{37} (\bibinfo{year}{2024}),
  \bibinfo{pages}{21999--22027}.
\newblock


\bibitem[Saad-Falcon et~al\mbox{.}(2024)]%
        {saad-falcon-etal-2024-ares}
\bibfield{author}{\bibinfo{person}{Jon Saad-Falcon}, \bibinfo{person}{Omar
  Khattab}, \bibinfo{person}{Christopher Potts}, {and} \bibinfo{person}{Matei
  Zaharia}.} \bibinfo{year}{2024}\natexlab{}.
\newblock \showarticletitle{{ARES}: An Automated Evaluation Framework for
  Retrieval-Augmented Generation Systems}. In
  \bibinfo{booktitle}{\emph{Proceedings of the Conference of the North American
  Chapter of the Association for Computational Linguistics: Human Language
  Technologies (Volume 1: Long Papers)}}. \bibinfo{publisher}{Association for
  Computational Linguistics}, \bibinfo{address}{Mexico City, Mexico},
  \bibinfo{pages}{338--354}.
\newblock
\href{https://doi.org/10.18653/v1/2024.naacl-long.20}{doi:\nolinkurl{10.18653/v1/2024.naacl-long.20}}


\bibitem[Salemi and Zamani(2024)]%
        {Salemi2024Evaluating}
\bibfield{author}{\bibinfo{person}{Alireza Salemi} {and} \bibinfo{person}{Hamed
  Zamani}.} \bibinfo{year}{2024}\natexlab{}.
\newblock \showarticletitle{Evaluating Retrieval Quality in Retrieval-Augmented
  Generation}. In \bibinfo{booktitle}{\emph{Proceedings of the International
  ACM SIGIR Conference on Research and Development in Information Retrieval}}.
  \bibinfo{publisher}{Association for Computing Machinery},
  \bibinfo{address}{New York, NY, USA}, \bibinfo{pages}{2395–2400}.
\newblock
\showISBNx{9798400704314}
\href{https://doi.org/10.1145/3626772.3657957}{doi:\nolinkurl{10.1145/3626772.3657957}}


\bibitem[Samarajeewa et~al\mbox{.}(2024)]%
        {Samarajeewa2024Causal}
\bibfield{author}{\bibinfo{person}{Chamod Samarajeewa}, \bibinfo{person}{Daswin
  De~Silva}, \bibinfo{person}{Evgeny Osipov}, \bibinfo{person}{Damminda
  Alahakoon}, {and} \bibinfo{person}{Milos Manic}.}
  \bibinfo{year}{2024}\natexlab{}.
\newblock \showarticletitle{Causal Reasoning in Large Language Models using
  Causal Graph Retrieval Augmented Generation}. In
  \bibinfo{booktitle}{\emph{Proceedings of the International Conference on
  Human System Interaction}}. \bibinfo{publisher}{IEEE},
  \bibinfo{address}{Paris, France}, \bibinfo{pages}{1--6}.
\newblock
\href{https://doi.org/10.1109/HSI61632.2024.10613566}{doi:\nolinkurl{10.1109/HSI61632.2024.10613566}}


\bibitem[Sarmah et~al\mbox{.}(2024)]%
        {Sarmah2024HybridRAG}
\bibfield{author}{\bibinfo{person}{Bhaskarjit Sarmah}, \bibinfo{person}{Dhagash
  Mehta}, \bibinfo{person}{Benika Hall}, \bibinfo{person}{Rohan Rao},
  \bibinfo{person}{Sunil Patel}, {and} \bibinfo{person}{Stefano Pasquali}.}
  \bibinfo{year}{2024}\natexlab{}.
\newblock \showarticletitle{HybridRAG: Integrating Knowledge Graphs and Vector
  Retrieval Augmented Generation for Efficient Information Extraction}. In
  \bibinfo{booktitle}{\emph{Proceedings of the ACM International Conference on
  AI in Finance}}. \bibinfo{publisher}{Association for Computing Machinery},
  \bibinfo{address}{New York, NY, USA}, \bibinfo{pages}{608–616}.
\newblock
\showISBNx{9798400710810}
\href{https://doi.org/10.1145/3677052.3698671}{doi:\nolinkurl{10.1145/3677052.3698671}}


\bibitem[Sawarkar et~al\mbox{.}(2024)]%
        {Sawarkar2024BlendedRAG}
\bibfield{author}{\bibinfo{person}{Kunal Sawarkar}, \bibinfo{person}{Abhilasha
  Mangal}, {and} \bibinfo{person}{Shivam~Raj Solanki}.}
  \bibinfo{year}{2024}\natexlab{}.
\newblock \showarticletitle{Blended RAG: Improving RAG (Retriever-Augmented
  Generation) Accuracy with Semantic Search and Hybrid Query-Based Retrievers}.
  In \bibinfo{booktitle}{\emph{Proceedings of IEEE International Conference on
  Multimedia Information Processing and Retrieval}}. \bibinfo{publisher}{IEEE},
  \bibinfo{address}{San Jose, CA, USA}, \bibinfo{pages}{155--161}.
\newblock
\href{https://doi.org/10.1109/MIPR62202.2024.00031}{doi:\nolinkurl{10.1109/MIPR62202.2024.00031}}


\bibitem[Seo et~al\mbox{.}(2024)]%
        {Seo2024Seo}
\bibfield{author}{\bibinfo{person}{JooYoung Seo}, \bibinfo{person}{Sanchita~S.
  Kamath}, \bibinfo{person}{Aziz Zeidieh}, \bibinfo{person}{Saairam Venkatesh},
  {and} \bibinfo{person}{Sean McCurry}.} \bibinfo{year}{2024}\natexlab{}.
\newblock \showarticletitle{MAIDR Meets AI: Exploring Multimodal LLM-Based Data
  Visualization Interpretation by and with Blind and Low-Vision Users}. In
  \bibinfo{booktitle}{\emph{Proceedings of the International ACM SIGACCESS
  Conference on Computers and Accessibility}}. \bibinfo{publisher}{Association
  for Computing Machinery}, \bibinfo{address}{New York, NY, USA},
  \bibinfo{pages}{1--31}.
\newblock
\showISBNx{9798400706776}
\href{https://doi.org/10.1145/3663548.3675660}{doi:\nolinkurl{10.1145/3663548.3675660}}


\bibitem[Shao et~al\mbox{.}(2024)]%
        {VEQA2023}
\bibfield{author}{\bibinfo{person}{Zekai Shao}, \bibinfo{person}{Shuran Sun},
  \bibinfo{person}{Yuheng Zhao}, \bibinfo{person}{Siyuan Wang},
  \bibinfo{person}{Zhongyu Wei}, \bibinfo{person}{Tao Gui},
  \bibinfo{person}{Cagatay Turkay}, {and} \bibinfo{person}{Siming Chen}.}
  \bibinfo{year}{2024}\natexlab{}.
\newblock \showarticletitle{Visual Explanation for Open-Domain Question
  Answering With BERT}.
\newblock \bibinfo{journal}{\emph{IEEE Transactions on Visualization and
  Computer Graphics}} \bibinfo{volume}{30}, \bibinfo{number}{7}
  (\bibinfo{year}{2024}), \bibinfo{pages}{3779--3797}.
\newblock
\href{https://doi.org/10.1109/TVCG.2023.3243676}{doi:\nolinkurl{10.1109/TVCG.2023.3243676}}


\bibitem[Sivasothy et~al\mbox{.}(2024)]%
        {sivasothy2024ragprobeautomatedapproachevaluating}
\bibfield{author}{\bibinfo{person}{Shangeetha Sivasothy},
  \bibinfo{person}{Scott Barnett}, \bibinfo{person}{Stefanus Kurniawan},
  \bibinfo{person}{Zafaryab Rasool}, {and} \bibinfo{person}{Rajesh Vasa}.}
  \bibinfo{year}{2024}\natexlab{}.
\newblock \bibinfo{title}{RAGProbe: An Automated Approach for Evaluating RAG
  Applications}.
\newblock
\showeprint[arxiv]{2409.19019}~[cs.CL]
\urldef\tempurl%
\url{https://arxiv.org/abs/2409.19019}
\showURL{%
\tempurl}


\bibitem[Song et~al\mbox{.}(2023)]%
        {Song2023chi}
\bibfield{author}{\bibinfo{person}{Da Song}, \bibinfo{person}{Zhijie Wang},
  \bibinfo{person}{Yuheng Huang}, \bibinfo{person}{Lei Ma}, {and}
  \bibinfo{person}{Tianyi Zhang}.} \bibinfo{year}{2023}\natexlab{}.
\newblock \showarticletitle{DeepLens: Interactive Out-of-distribution Data
  Detection in NLP Models}. In \bibinfo{booktitle}{\emph{Proceedings of the CHI
  Conference on Human Factors in Computing Systems}}.
  \bibinfo{publisher}{Association for Computing Machinery},
  \bibinfo{address}{New York, NY, USA}, \bibinfo{pages}{1--17}.
\newblock
\showISBNx{9781450394215}
\href{https://doi.org/10.1145/3544548.3580741}{doi:\nolinkurl{10.1145/3544548.3580741}}


\bibitem[Soudani et~al\mbox{.}(2024)]%
        {Soudani2024FineTuning}
\bibfield{author}{\bibinfo{person}{Heydar Soudani}, \bibinfo{person}{Evangelos
  Kanoulas}, {and} \bibinfo{person}{Faegheh Hasibi}.}
  \bibinfo{year}{2024}\natexlab{}.
\newblock \showarticletitle{Fine Tuning vs. Retrieval Augmented Generation for
  Less Popular Knowledge}. In \bibinfo{booktitle}{\emph{Proceedings of the
  Annual International ACM SIGIR Conference on Research and Development in
  Information Retrieval in the Asia Pacific Region}}.
  \bibinfo{publisher}{Association for Computing Machinery},
  \bibinfo{address}{New York, NY, USA}, \bibinfo{pages}{12–--22}.
\newblock
\showISBNx{9798400707247}
\href{https://doi.org/10.1145/3673791.3698415}{doi:\nolinkurl{10.1145/3673791.3698415}}


\bibitem[Susnjak et~al\mbox{.}(2025)]%
        {Susnjak2025Automating}
\bibfield{author}{\bibinfo{person}{Teo Susnjak}, \bibinfo{person}{Peter Hwang},
  \bibinfo{person}{Napoleon Reyes}, \bibinfo{person}{Andre L.~C. Barczak},
  \bibinfo{person}{Timothy McIntosh}, {and} \bibinfo{person}{Surangika
  Ranathunga}.} \bibinfo{year}{2025}\natexlab{}.
\newblock \showarticletitle{Automating Research Synthesis with Domain-Specific
  Large Language Model Fine-Tuning}.
\newblock \bibinfo{journal}{\emph{ACM Transactions on Knowledge Discovery from
  Data}} \bibinfo{volume}{19}, \bibinfo{number}{3} (\bibinfo{year}{2025}),
  \bibinfo{pages}{1--–39}.
\newblock
\showISSN{1556-4681}
\href{https://doi.org/10.1145/3715964}{doi:\nolinkurl{10.1145/3715964}}


\bibitem[Tailor et~al\mbox{.}(2024)]%
        {Prashant2024Comparative}
\bibfield{author}{\bibinfo{person}{Prashant~D. Tailor},
  \bibinfo{person}{Lauren~A. Dalvin}, \bibinfo{person}{John~J. Chen},
  \bibinfo{person}{Raymond Iezzi}, \bibinfo{person}{Timothy~W. Olsen},
  \bibinfo{person}{Brittni~A. Scruggs}, \bibinfo{person}{Andrew~J. Barkmeier},
  \bibinfo{person}{Sophie~J. Bakri}, \bibinfo{person}{Edwin~H. Ryan},
  \bibinfo{person}{Peter~H. Tang}, \bibinfo{person}{D.~Wilkin. Parke},
  \bibinfo{person}{Peter~J. Belin}, \bibinfo{person}{Jayanth Sridhar},
  \bibinfo{person}{David Xu}, \bibinfo{person}{Ajay~E. Kuriyan},
  \bibinfo{person}{Yoshihiro Yonekawa}, {and} \bibinfo{person}{Matthew~R.
  Starr}.} \bibinfo{year}{2024}\natexlab{}.
\newblock \showarticletitle{A Comparative Study of Responses to Retina
  Questions from Either Experts, Expert-Edited Large Language Models, or
  Expert-Edited Large Language Models Alone}.
\newblock \bibinfo{journal}{\emph{Ophthalmology Science}} \bibinfo{volume}{4},
  \bibinfo{number}{4} (\bibinfo{year}{2024}), \bibinfo{pages}{100485}.
\newblock
\showISSN{2666-9145}
\href{https://doi.org/10.1016/j.xops.2024.100485}{doi:\nolinkurl{10.1016/j.xops.2024.100485}}


\bibitem[van Schaik and Pugh(2024)]%
        {van2024Field}
\bibfield{author}{\bibinfo{person}{Tempest~A. van Schaik} {and}
  \bibinfo{person}{Brittany Pugh}.} \bibinfo{year}{2024}\natexlab{}.
\newblock \showarticletitle{A Field Guide to Automatic Evaluation of
  LLM-Generated Summaries}. In \bibinfo{booktitle}{\emph{Proceedings of the
  International ACM SIGIR Conference on Research and Development in Information
  Retrieval}}. \bibinfo{publisher}{Association for Computing Machinery},
  \bibinfo{address}{New York, NY, USA}, \bibinfo{pages}{2832--2836}.
\newblock
\showISBNx{9798400704314}
\href{https://doi.org/10.1145/3626772.3661346}{doi:\nolinkurl{10.1145/3626772.3661346}}


\bibitem[Wang et~al\mbox{.}(2024)]%
        {wang2023commonsensevis}
\bibfield{author}{\bibinfo{person}{Xingbo Wang}, \bibinfo{person}{Renfei
  Huang}, \bibinfo{person}{Zhihua Jin}, \bibinfo{person}{Tianqing Fang}, {and}
  \bibinfo{person}{Huamin Qu}.} \bibinfo{year}{2024}\natexlab{}.
\newblock \showarticletitle{CommonsenseVIS: Visualizing and Understanding
  Commonsense Reasoning Capabilities of Natural Language Models}.
\newblock \bibinfo{journal}{\emph{IEEE Transactions on Visualization and
  Computer Graphics}} \bibinfo{volume}{30}, \bibinfo{number}{1}
  (\bibinfo{year}{2024}), \bibinfo{pages}{273--283}.
\newblock
\href{https://doi.org/10.1109/TVCG.2023.3327153}{doi:\nolinkurl{10.1109/TVCG.2023.3327153}}


\bibitem[Wu et~al\mbox{.}(2024)]%
        {Wu2024Survey}
\bibfield{author}{\bibinfo{person}{L. Wu}, \bibinfo{person}{Z. Zheng},
  \bibinfo{person}{Z. Qiu}, {et~al\mbox{.}}} \bibinfo{year}{2024}\natexlab{}.
\newblock \showarticletitle{A survey on large language models for
  recommendation}.
\newblock \bibinfo{journal}{\emph{World Wide Web}}  \bibinfo{volume}{27}
  (\bibinfo{year}{2024}).
\newblock
\href{https://doi.org/10.1007/s11280-024-01291-2}{doi:\nolinkurl{10.1007/s11280-024-01291-2}}


\bibitem[Xie et~al\mbox{.}(2024)]%
        {Xie2024uist}
\bibfield{author}{\bibinfo{person}{Liwenhan Xie}, \bibinfo{person}{Chengbo
  Zheng}, \bibinfo{person}{Haijun Xia}, \bibinfo{person}{Huamin Qu}, {and}
  \bibinfo{person}{Chen Zhu-Tian}.} \bibinfo{year}{2024}\natexlab{}.
\newblock \showarticletitle{WaitGPT: Monitoring and Steering Conversational LLM
  Agent in Data Analysis with On-the-Fly Code Visualization}. In
  \bibinfo{booktitle}{\emph{Proceedings of the Annual ACM Symposium on User
  Interface Software and Technology}}. \bibinfo{publisher}{Association for
  Computing Machinery}, \bibinfo{address}{New York, NY, USA},
  \bibinfo{pages}{1--14}.
\newblock
\showISBNx{9798400706288}
\href{https://doi.org/10.1145/3654777.3676374}{doi:\nolinkurl{10.1145/3654777.3676374}}


\bibitem[Yan et~al\mbox{.}(2025)]%
        {yan2024knownet}
\bibfield{author}{\bibinfo{person}{Youfu Yan}, \bibinfo{person}{Yu Hou},
  \bibinfo{person}{Yongkang Xiao}, \bibinfo{person}{Rui Zhang}, {and}
  \bibinfo{person}{Qianwen Wang}.} \bibinfo{year}{2025}\natexlab{}.
\newblock \showarticletitle{KNowNEt:Guided Health Information Seeking from LLMs
  via Knowledge Graph Integration}.
\newblock \bibinfo{journal}{\emph{IEEE Transactions on Visualization and
  Computer Graphics}} \bibinfo{volume}{31}, \bibinfo{number}{1}
  (\bibinfo{year}{2025}), \bibinfo{pages}{547--557}.
\newblock
\href{https://doi.org/10.1109/TVCG.2024.3456364}{doi:\nolinkurl{10.1109/TVCG.2024.3456364}}


\bibitem[Yang et~al\mbox{.}(2024a)]%
        {yang2024harnessing}
\bibfield{author}{\bibinfo{person}{Jingfeng Yang}, \bibinfo{person}{Hongye
  Jin}, \bibinfo{person}{Ruixiang Tang}, \bibinfo{person}{Xiaotian Han},
  \bibinfo{person}{Qizhang Feng}, \bibinfo{person}{Haoming Jiang},
  \bibinfo{person}{Shaochen Zhong}, \bibinfo{person}{Bing Yin}, {and}
  \bibinfo{person}{Xia Hu}.} \bibinfo{year}{2024}\natexlab{a}.
\newblock \showarticletitle{Harnessing the Power of LLMs in Practice: A Survey
  on ChatGPT and Beyond}.
\newblock \bibinfo{journal}{\emph{ACM Transactions on Knowledge Discovery from
  Data}} \bibinfo{volume}{18}, \bibinfo{number}{6} (\bibinfo{year}{2024}),
  \bibinfo{pages}{1--32}.
\newblock
\showISSN{1556-4681}
\href{https://doi.org/10.1145/3649506}{doi:\nolinkurl{10.1145/3649506}}


\bibitem[Yang et~al\mbox{.}(2024b)]%
        {Yang2024Advances}
\bibfield{author}{\bibinfo{person}{Xiao Yang}, \bibinfo{person}{Kai Sun},
  \bibinfo{person}{Hao Xin}, \bibinfo{person}{Yushi Sun},
  \bibinfo{person}{Nikita Bhalla}, \bibinfo{person}{Xiangsen Chen},
  \bibinfo{person}{Sajal Choudhary}, \bibinfo{person}{Rongze~Daniel Gui},
  \bibinfo{person}{Ziran~Will Jiang}, \bibinfo{person}{Ziyu Jiang},
  \bibinfo{person}{Lingkun Kong}, \bibinfo{person}{Brian Moran},
  \bibinfo{person}{Jiaqi Wang}, \bibinfo{person}{Yifan~Ethan Xu},
  \bibinfo{person}{An Yan}, \bibinfo{person}{Chenyu Yang},
  \bibinfo{person}{Eting Yuan}, \bibinfo{person}{Hanwen Zha},
  \bibinfo{person}{Nan Tang}, \bibinfo{person}{Lei Chen},
  \bibinfo{person}{Nicolas Scheffer}, \bibinfo{person}{Yue Liu},
  \bibinfo{person}{Nirav Shah}, \bibinfo{person}{Rakesh Wanga},
  \bibinfo{person}{Anuj Kumar}, \bibinfo{person}{Wen-tau Yih}, {and}
  \bibinfo{person}{Xin~Luna Dong}.} \bibinfo{year}{2024}\natexlab{b}.
\newblock \showarticletitle{CRAG - Comprehensive RAG Benchmark}. In
  \bibinfo{booktitle}{\emph{Proceedings of the Conference on Neural Information
  Processing Systems, Track on Datasets and Benchmarks}}.
  \bibinfo{pages}{10470--10490}.
\newblock
\href{https://doi.org/10.48550/arXiv.2406.04744}{doi:\nolinkurl{10.48550/arXiv.2406.04744}}


\bibitem[Yu et~al\mbox{.}(2024)]%
        {yu2024evaluation}
\bibfield{author}{\bibinfo{person}{Hao Yu}, \bibinfo{person}{Aoran Gan},
  \bibinfo{person}{Kai Zhang}, \bibinfo{person}{Shiwei Tong},
  \bibinfo{person}{Qi Liu}, {and} \bibinfo{person}{Zhaofeng Liu}.}
  \bibinfo{year}{2024}\natexlab{}.
\newblock \showarticletitle{Evaluation of retrieval-augmented generation: A
  survey}. In \bibinfo{booktitle}{\emph{Proceedings of CCF Conference on Big
  Data}}. \bibinfo{publisher}{Springer}, \bibinfo{address}{Singapore},
  \bibinfo{pages}{102--120}.
\newblock
\href{https://doi.org/10.1007/978-981-96-1024-2_8}{doi:\nolinkurl{10.1007/978-981-96-1024-2_8}}


\bibitem[Yue et~al\mbox{.}(2023)]%
        {yue-etal-2023-automatic}
\bibfield{author}{\bibinfo{person}{Xiang Yue}, \bibinfo{person}{Boshi Wang},
  \bibinfo{person}{Ziru Chen}, \bibinfo{person}{Kai Zhang}, \bibinfo{person}{Yu
  Su}, {and} \bibinfo{person}{Huan Sun}.} \bibinfo{year}{2023}\natexlab{}.
\newblock \showarticletitle{Automatic Evaluation of Attribution by Large
  Language Models}. In \bibinfo{booktitle}{\emph{Findings of the Association
  for Computational Linguistics: EMNLP}}. \bibinfo{publisher}{Association for
  Computational Linguistics}, \bibinfo{address}{Singapore},
  \bibinfo{pages}{4615--4635}.
\newblock
\href{https://doi.org/10.18653/v1/2023.findings-emnlp.307}{doi:\nolinkurl{10.18653/v1/2023.findings-emnlp.307}}


\bibitem[Zamani et~al\mbox{.}(2022)]%
        {Zamani2022Retrieval}
\bibfield{author}{\bibinfo{person}{Hamed Zamani}, \bibinfo{person}{Fernando
  Diaz}, \bibinfo{person}{Mostafa Dehghani}, \bibinfo{person}{Donald Metzler},
  {and} \bibinfo{person}{Michael Bendersky}.} \bibinfo{year}{2022}\natexlab{}.
\newblock \showarticletitle{Retrieval-Enhanced Machine Learning}. In
  \bibinfo{booktitle}{\emph{Proceedings of the 45th International ACM SIGIR
  Conference on Research and Development in Information Retrieval}}.
  \bibinfo{publisher}{Association for Computing Machinery},
  \bibinfo{address}{New York, NY, USA}, \bibinfo{pages}{2875–2886}.
\newblock
\showISBNx{9781450387323}
\href{https://doi.org/10.1145/3477495.3531722}{doi:\nolinkurl{10.1145/3477495.3531722}}


\bibitem[Zhang et~al\mbox{.}(2024)]%
        {zhang-etal-2024-clamber}
\bibfield{author}{\bibinfo{person}{Tong Zhang}, \bibinfo{person}{Peixin Qin},
  \bibinfo{person}{Yang Deng}, \bibinfo{person}{Chen Huang},
  \bibinfo{person}{Wenqiang Lei}, \bibinfo{person}{Junhong Liu},
  \bibinfo{person}{Dingnan Jin}, \bibinfo{person}{Hongru Liang}, {and}
  \bibinfo{person}{Tat-Seng Chua}.} \bibinfo{year}{2024}\natexlab{}.
\newblock \showarticletitle{{CLAMBER}: A Benchmark of Identifying and
  Clarifying Ambiguous Information Needs in Large Language Models}. In
  \bibinfo{booktitle}{\emph{Proceedings of the Annual Meeting of the
  Association for Computational Linguistics}}. \bibinfo{publisher}{Association
  for Computational Linguistics}, \bibinfo{address}{Bangkok, Thailand},
  \bibinfo{pages}{10746--10766}.
\newblock
\href{https://doi.org/10.18653/v1/2024.acl-long.578}{doi:\nolinkurl{10.18653/v1/2024.acl-long.578}}


\bibitem[Zhao et~al\mbox{.}(2024a)]%
        {Zhao2024Explainability}
\bibfield{author}{\bibinfo{person}{Haiyan Zhao}, \bibinfo{person}{Hanjie Chen},
  \bibinfo{person}{Fan Yang}, \bibinfo{person}{Ninghao Liu},
  \bibinfo{person}{Huiqi Deng}, \bibinfo{person}{Hengyi Cai},
  \bibinfo{person}{Shuaiqiang Wang}, \bibinfo{person}{Dawei Yin}, {and}
  \bibinfo{person}{Mengnan Du}.} \bibinfo{year}{2024}\natexlab{a}.
\newblock \showarticletitle{Explainability for Large Language Models: A
  Survey}.
\newblock \bibinfo{journal}{\emph{ACM Transactions on Intelligent Systems and
  Technology}} \bibinfo{volume}{15}, \bibinfo{number}{2}
  (\bibinfo{year}{2024}), \bibinfo{pages}{1–38}.
\newblock
\showISSN{2157-6904}
\href{https://doi.org/10.1145/3639372}{doi:\nolinkurl{10.1145/3639372}}


\bibitem[Zhao et~al\mbox{.}(2024b)]%
        {Zhao2024Dense}
\bibfield{author}{\bibinfo{person}{Wayne~Xin Zhao}, \bibinfo{person}{Jing Liu},
  \bibinfo{person}{Ruiyang Ren}, {and} \bibinfo{person}{Ji-Rong Wen}.}
  \bibinfo{year}{2024}\natexlab{b}.
\newblock \showarticletitle{Dense Text Retrieval Based on Pretrained Language
  Models: A Survey}.
\newblock \bibinfo{journal}{\emph{ACM Transactions on Information Systems}}
  \bibinfo{volume}{42}, \bibinfo{number}{4} (\bibinfo{year}{2024}),
  \bibinfo{pages}{1--60}.
\newblock
\showISSN{1046-8188}
\href{https://doi.org/10.1145/3637870}{doi:\nolinkurl{10.1145/3637870}}


\bibitem[Zhuo et~al\mbox{.}(2024)]%
        {zhuo-etal-2024-prosa}
\bibfield{author}{\bibinfo{person}{Jingming Zhuo}, \bibinfo{person}{Songyang
  Zhang}, \bibinfo{person}{Xinyu Fang}, \bibinfo{person}{Haodong Duan},
  \bibinfo{person}{Dahua Lin}, {and} \bibinfo{person}{Kai Chen}.}
  \bibinfo{year}{2024}\natexlab{}.
\newblock \showarticletitle{{P}ro{SA}: Assessing and Understanding the Prompt
  Sensitivity of {LLM}s}. In \bibinfo{booktitle}{\emph{Findings of the
  Association for Computational Linguistics: EMNLP}}.
  \bibinfo{publisher}{Association for Computational Linguistics},
  \bibinfo{address}{Miami, Florida, USA}, \bibinfo{pages}{1950--1976}.
\newblock
\href{https://doi.org/10.18653/v1/2024.findings-emnlp.108}{doi:\nolinkurl{10.18653/v1/2024.findings-emnlp.108}}


\bibitem[Şakar and Emekci(2025)]%
        {Sakar2025rag}
\bibfield{author}{\bibinfo{person}{T. Şakar} {and} \bibinfo{person}{H.
  Emekci}.} \bibinfo{year}{2025}\natexlab{}.
\newblock \showarticletitle{Maximizing {RAG} Efficiency: A Comparative Analysis
  of {RAG} Methods}.
\newblock \bibinfo{journal}{\emph{Natural Language Processing}}
  \bibinfo{volume}{31}, \bibinfo{number}{1} (\bibinfo{year}{2025}),
  \bibinfo{pages}{1--25}.
\newblock
\href{https://doi.org/10.1017/nlp.2024.53}{doi:\nolinkurl{10.1017/nlp.2024.53}}


\end{thebibliography}

\clearpage

\appendix

\section{Research Methods}
\label{appendix:A}
\subsection{Formative Study Participants}

In our formative study, we recruited 12 participants (E1-E12) with varying degrees of experience with large language models and from diverse research backgrounds. Table \ref{tab:formative_participants} presents the demographic information and expertise details of these participants.

\begin{table}[ht]
\centering
\caption{Demographic Information and Background of Formative Study Participants}
\label{tab:formative_participants}
\begin{tabular}{cccp{2.8cm}}
\toprule
\textbf{ID} & \textbf{Gender} & \textbf{LLM Experience} & \textbf{Research Background} \\
\midrule
E1 & Male & 2 years & RAG and LLM Agents \\
E2 & Male & 3 years & RAG Researcher \\
E3 & Male & 2 years & RAG \\
E4 & Male & 3 years & RAG Specialist \\
E5 & Male & 3 years & Nuclear Physics \\
E6 & Female & 1 year & Public Health \\
E7 & Female & 1 year & Business Analysis \\
E8 & Male & 2 years & Backend Development \\
E9 & Male & 1 year & Quantum Computing \\
E10 & Male & 2 years & Medical Science \\
E11 & Male & 3 years & Computer Science \\
E12 & Female & 1 year & Artificial Intelligence \\
\bottomrule
\end{tabular}
\end{table}

The participants were recruited through university mailing lists and professional networks. They all had prior experience with large language models, ranging from 1 to 3 years. Their diverse research backgrounds allowed us to gather insights from multiple perspectives, enhancing the comprehensiveness of our formative study.

\subsection{User Study Participants}

For our user study, we recruited 11 participants (P1-P11) to evaluate the effectiveness and usability of \sysname{}. Table \ref{tab:user_participants} presents the demographic information and background details of these participants.

\begin{table}[ht]
\centering
\caption{Demographic Information and Background of User Study Participants}
\label{tab:user_participants}
\begin{tabular}{cccp{2.8cm}}
\toprule
\textbf{ID} & \textbf{Gender} & \textbf{LLM Experience} & \textbf{Research Background} \\
\midrule
P1 & Male & 3 years & RAG Researcher \\
P2 & Male & 2 years & RAG  \\
P3 & Male & 2 years & RAG  \\
P4 & Male & 2 years & LLM Agents \\
P5 & Male & 2 years & HCI  \\
P6 & Male & 3 years & Computer Science  \\
P7 & Female & 2 years & HCI  \\
P8 & Female & 1 year & Public Health  \\
P9 & Male & 1 year & Quantum Computing \\
P10 & Male & 3 years & Nuclear Physics \\
P11 & Female & 2 years & HCI \\
\bottomrule
\end{tabular}
\end{table}

The user study participants were selected to represent a range of expertise levels in large language models and diverse academic backgrounds. Notably, some participants (P1/E2, P2/E3, P4/E1, P6/E11, P8/E6, P9/E9, P10/E5) also participated in our formative study, providing valuable continuity in our research process. This overlap allowed these participants to compare their experiences with and without the \sysname{} system.

\section{Technical Implementation Details}

\label{appendix:B}

This appendix details the technical infrastructure and methodologies supporting our \sysname{} system.

\paragraph{Model Infrastructure.} Our system employs Llama3-70B-4bit as the primary language model, deployed locally across four NVIDIA GeForce RTX 3090 GPUs in a distributed inference configuration.

\paragraph{Confidence Score Acquisition.} Confidence scores are obtained through two distinct pathways: (1) For open-source models, confidence values are extracted directly through local model interfaces that provide access to internal probability distributions; (2) For proprietary models, confidence scores are retrieved via OpenAI API endpoints that expose model uncertainty metrics.

\paragraph{Relevance Node Classification.} The determination of node relevance employs a multi-stage classification pipeline. Candidate nodes are evaluated against ground truth annotations using semantic similarity measures. Nodes deemed relevant are highlighted in blue or grey within the visualization interface. We implements a multi-step pipeline as follows:

\begin{verbatim}
Algorithm: Named Entity Processing Pipeline
1. EXTRACT entities from ground truth using NLTK
2. FOR each entity:
   a. CHECK cache for existing synonyms/antonyms
   b. IF cached: RETURN cached results
   c. ELSE:
      - GENERATE contextual synonyms/antonyms via LLM
      - CACHE results in database
3. COMPARE entity sets between ground truth and model response
4. COMPUTE semantic similarity scores
5. RETURN evaluation metrics (precision, recall, F1) 
to determine color
\end{verbatim}

The pipeline first extracts named entities from ground truth documents using NLTK's pre-trained named entity recognition model. These entities undergo contextual filtering through a language model (DeepSeek-V3 in our experiments) that evaluates contextual relevance and generates semantically related terms, including synonyms and antonyms. The expanded entity vocabulary is then systematically compared against model-generated responses to identify semantic alignments and discrepancies.

\end{document}